\def\figwidth{0.75\linewidth}

\def\keys{Bayesian forecasting $|$ Public health situation awareness
  $|$ Data-driven epidemics $|$ Compartment-based model $|$ Kalman
  filtering}

\documentclass[9pt,lineno,a4paper]{article}
\usepackage{authblk}
\usepackage[cm]{fullpage}
\usepackage{amsmath,amsfonts,amssymb,amsthm}
\usepackage[english]{babel}
\usepackage[utf8]{inputenc}
\usepackage{enumerate}
\usepackage{graphicx}
\usepackage{graphicx}
\usepackage{color}
\usepackage{sidecap}
\usepackage[pdftitle={Bayesian Monitoring of COVID-19 in Sweden},
  pdfauthor={Marin, Runvik, Medvedev, Engblom},
  pdffitwindow=true,
  breaklinks=true,
  colorlinks=true,
  urlcolor=blue,
  linkcolor=red,
  citecolor=red,
  anchorcolor=red]{hyperref}
\usepackage[numbers,sort&compress]{natbib}
\usepackage{lineno}                
\pagewiselinenumbers            
\nolinenumbers 

\usepackage[symbol]{footmisc}

\author[a]{Robin Marin}
\author[b]{Håkan Runvik}
\author[b]{Alexander Medvedev}
\author[a]{Stefan Engblom\thanks{Corresponding author, telephone
    +46-18-471 27 54, fax +46-18-51 19 25.}}

\affil[a]{{\footnotesize Division of Scientific Computing, Department
    of Information Technology, Uppsala University, SE-751 05 Uppsala,
    Sweden. E-mail: \href{mailto:robin.marin@it.uu.se}{robin.marin},
    \href{mailto:stefane@it.uu.se}{stefane@it.uu.se}.}}

\affil[a]{{\footnotesize Division of Systems and Control, Department
    of Information Technology, Uppsala University, SE-751 05 Uppsala,
    Sweden. E-mail: \href{mailto:hakan.runvik@it.uu.se}{hakan.runvik},
    \href{mailto:alexander.medvedev@it.uu.se}{alexander.medvedev@it.uu.se}.}}

\date{\today}

\usepackage{gensymb}
\usepackage{booktabs}
\usepackage{xr}

\usepackage{subcaption}
\usepackage{soul}

\newcommand{\review}[1]{#1}
\newcommand{\reviewb}[1]{#1}

\newcommand{\CI}[5]{\numprint{#3} [\numprint{#2},\,\numprint{#4}]} 
\newcommand{\CIedge}[5]{[\numprint{#1},\,\numprint{#5}]} 

\renewcommand{\Pr}{\mathbf{P}}
\DeclareMathOperator{\Expect}{\mathbb{E}}
\DeclareMathOperator*{\argmin}{arg\,min}
\DeclareMathOperator{\diam}{\mbox{diam}}

\newcommand{\matmet}{Material and Methods}

\usepackage{numprint}
\npthousandsep{\,}
\npdecimalsign{.}

\title{Bayesian Monitoring of COVID-19 in Sweden}

\begin{document}

\maketitle

\begin{abstract}


  %
  %
  In an effort to provide regional decision support for the public
  healthcare, we design a data-driven compartment-based model of
  COVID-19 in Sweden. From national hospital statistics we derive
  parameter priors, and we develop linear filtering techniques to
  drive the simulations given data in the form of daily healthcare
  demands. We additionally propose a posterior marginal estimator
  which \review{provides for an improved temporal} resolution of the
  reproduction number estimate \review{as well as supports robustness
    checks via} a parametric bootstrap procedure.

  From our computational approach we obtain a Bayesian model of
  predictive value which provides important insight into the
  progression of the disease, including estimates of the effective
  reproduction number, the infection fatality rate, and the
  regional-level immunity. We successfully validate our posterior
  model against several different sources, including outputs from
  extensive screening programs. Since our required data in comparison
  is easy and non-sensitive to collect, we argue that our approach is
  particularly promising as a tool to support monitoring and decisions
  within public health.

  \medskip \par
  \noindent
  \textbf{Significance:} 
  Using public data from Swedish patient registries we develop a
  national-scale computational model of COVID-19. The parametrized
  model produces valuable weekly predictions of healthcare demands at
  the regional level and validates well against several different
  sources. We also obtain critical epidemiological insights into the
  disease progression, including, e.g., reproduction number, immunity
  and disease fatality estimates. The success of the model hinges on
  our novel use of filtering techniques which allows us to design an
  accurate data-driven procedure using data exclusively from
  healthcare demands, i.e., our approach does not rely on public
  testing and is therefore very cost-effective.

\end{abstract}


\section*{Introduction}



The results in this paper stem from the work carried out within the
cross-disciplinary research project CRUSH Covid at Uppsala
University\footnote[2]{\url{https://www.uu.se/forskning/projekt/crush-covid/}}.
Starting in the fall 2020, every week the group published a widely
circulated report covering the region’s COVID status by, e.g.,
collecting data from PCR tests, mobile apps, wastewater analysis, and
health care. Our contribution consisted of a Bayesian disease-spread
model which provided decision support in the form of predictions of
health care demands as well as additional epidemiological insight.

There has been \review{a multitude of} attempts to model and forecast
the spread of the virus. \reviewb{A problem often encountered is that,
  although data might appear abundant, fitting a given model to large
  volumes of data of various quality does not necessarily imply a high
  prediction accuracy} \cite{shinde2020forecasting}. \review{A related
  issue is to identify the right level of model granularity:} several
aspects of the disease transmission are \review{relevant} and need to
be modeled, from small scale \textit{in vitro} properties to global
interventions. Some modeling efforts \review{therefore} include
multiple levels of resolutions to capture, e.g., global travel
patterns \cite{davis2021cryptic} or local within-country dynamics
\cite{gatto2020spread}. Understanding how to combine the various
scales can substantially benefit the fidelity of scenario generators
\cite{jordan2021optimization}. It is fair to say that models which
have been in actual use are understudied due to time constraints and
therefore often lack a thorough uncertainty analysis
\cite{edeling2021impact, soltesz2020effect}. With the frequent lack of
high-quality data for the current state of the disease,
\emph{nowcasting} has been increasingly critical in decision making
\cite{wu2020nowcasting, altmejd2020nowcasting, hawryluk2021gaussian,
  wu2021nowcasting}.

\review{Since the situation concerns modeling under data limitations
  and with potentially large process uncertainties our proposed
  solution consists of a Bayesian framework. This approach} involves
adapting a linear noise approximation \cite{fearnhead2014inference}
which enables fusion of different data sources and supports a
computationally cheap approximate likelihood function via linear
filters. We also investigate the posterior model not only through the
posterior predictive distribution \cite{gelman1995bayesian,
  gabry2019visualization}, but also by estimating the bias introduced
by the approximate likelihood using ideas from parametric bootstrap
\cite{engblom2020bayesian}.

Our disease spread model attempts to balance three key qualities:
interpretability, quantifiable uncertainty, and forecasting
accuracy. The posterior model was investigated through marginal
estimators, by comparing our results to several other sources, and by
bias estimates obtained via bootstrap arguments. As this paper
demonstrates, the achieved accuracy and robustness are quite
remarkable considering that no data from screening programs were used.


\section*{Material and Methods}



\review{Below we first summarize the Swedish publicly available data,
  and then present the associated design decisions made in developing
  the computational model. An important technical contribution lies in
  the techniques which support a computationally efficient approximate
  likelihood via linear filters.} Two `bootstrap' procedures are also
outlined: one for improving the temporal resolution of the
reproduction number estimates, and one for bounding the inversion bias
through the generation of synthetic data. Further technical details
concerning the derivation of the linear filters, data pre-processing,
and optimization algorithms are found in the Supporting Information
(SI).

\subsection*{Swedish COVID-19 data}



In Sweden, the publicly available \review{time series} data for the
COVID-19 pandemic fall in one of two categories: hospital load and
results from PCR testing. Cumulative national-level disease severity
statistics \review{have also been made available and} updated
approximately once a month throughout the pandemic. The 21 regional
councils compile hospital data and report the number of patients
undergoing inpatient or intensive care, and also the number of
deceased individuals. These numbers are reported on a daily basis and
have been judged to be of consistent quality over sufficiently long
periods of time to be used in our modeling. We retrieve those data
from the portal initiative \verb;c19.se;, which in turn collects the
data from the regional councils. For validation, we have compared with
official public registries, including the Swedish Public Health Agency
(PHA), the National Board of Health and Welfare (NBHW), and the
Swedish Intensive Care Registry (SIR). There are occasional
inconsistencies in the data which need to be filtered away; see the SI
for our quite basic approach for this.

The Swedish daily incidence as reported from PCR testing has been poor
in several periods of time due to restrictions and changes in testing
recommendations \cite{winblad2021soft}. In June 2020, the Swedish
government appointed a commission to evaluate COVID-19 measures,
including, among other things, the testing programs. They found that
the time from booking a PCR test to receiving the test results
exceeded six days across several regions during the period of time
studied in this paper, with additional time delays in
\review{publishing incidence results} on the regional and municipality
levels \cite{almgren2021kartlaggning}. For these reasons, we judged
the incidence data to be unreliable and excluded it from our
model. Note, however, the direct comparison with the incidence data
in Fig.~\ref{fig:Iinc_uppsala}.

Self-reporting via mobile apps has been proposed as a cheap and fast
alternative to PCR testing \cite{kennedy2022app}. However, the
validity of the signal depends on symptoms that overlap with other
respiratory infections. For example, the PHA noted a high occurrence
of symptoms of acute respiratory infections by the start of the
Swedish second wave in fall 2020 \cite{kennedy2022app,
  fhm2021vecka39}. Laboratory analyses of respiratory viruses later
indicated a high incidence of common colds caused by rhinoviruses
during the same period \cite{ki2022luftvag}. Another alternative data
source is the surveillance of wastewater \cite{saguti2021surveillance,
  galani2022sars}. Due to the signal's large relative noise ratio, we
did not consider it in this study but left it for future work.

\subsection*{Bayesian COVID-19 model}



Given the previous considerations of data sources, we formulate our
model around data in the form of daily observations of patients under
hospital care ($H$), intensive care ($W$), and reported deaths ($D$),
all in the 21 Swedish regions (counties). Any other sources of data
have been used for comparisons \textit{a posteriori} only.

In the standard SEIR model, susceptible individuals $S$ become exposed
$E$ (without symptoms), and after progressing to a symptomatic
infectious state $I$, they become recovered $R$. Based on the
available data, we extend the SEIR model with the states $(H,W,D)$,
and regard them as worsened states of the symptomatic infection such
that only a certain fraction of the \review{infected} individuals will
enter them. It is widely accepted that not all exposed
\review{individuals} become symptomatic \cite{byrne2020inferred} and
hence we also extend the model by including an asymptomatic state $A$
with no or very mild symptoms.

The transmission is usually driven by random or time-dependent contact
intensities between the susceptible and the infected individuals. As
in \cite{FreeLivingInfectiveStages, widgren2018spatio,
  engblom2020bayesian, VandenDriessche2017}, we rather consider
\emph{implicit spread} via an infectious pressure compartment
$\varphi$, an environmental state variable which models the current
force of infection of the virus \review{and which decays exponentially
  with time. \reviewb{Given that the decay rate of the infectious
    pressure is comparably fast and since direct spread can be
    understood as a timescale separation limit of indirect spread,
    most results should be robust with respect to this particular
    modeling choice (see also \cite{Benson2021}).}}

\review{The COVID-19 model in \cite{Keeling2021} sources the (explicit
  spread) infectivity from both symptomatic and asymptomatic
  carriers. We additionally allow for pre-symptomatic spread by
  sourcing the infectious pressure from all individuals in the states
  $(E,A,I)$. \reviewb{Originally, we intended for our framework to
    regularly simulate spread over a national network defined by
    commuting intensities between regions. However, in the end we did
    not routinely incorporate network spread in the model} since with
  our available data, letting the regions function as independent
  nodes reduces the computational complexity while giving a very
  similar data fit.}

\begin{SCfigure}[\sidecaptionrelwidth][tbhp]
  \centering
  \includegraphics[width=0.5\linewidth]
  {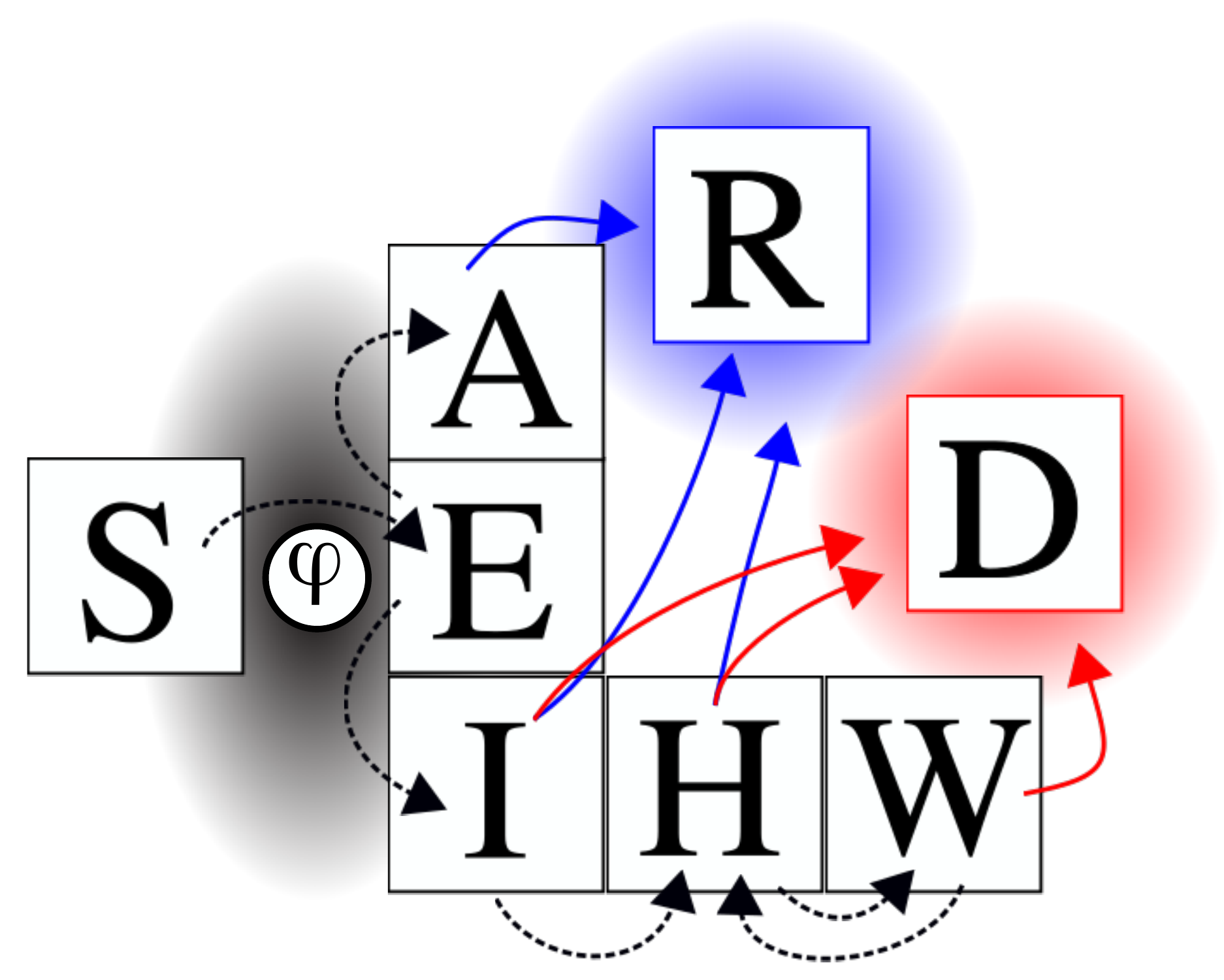}
  \caption{Proposed COVID-19 compartment model \reviewb{with indirect
      transmission}. (S)usceptible contract the virus by exposure to
    an infectious pressure ($\varphi$), become (E)xposed and then
    enter either the (A)symptomatic or the (I)nfected
    state. Symptomatic severity follows: (I)nfected to (H)ospitalized
    to (W)orsen (intensive care), and to (D)eath. Only states A, I,
    and H can (R)ecover from the disease. The observables are H, W,
    and D. \reviewb{The infectious pressure is sourced from individuals
      in the states $(E,A,I)$ and decays exponentially with time.}}
  \label{fig:compartments}
\end{SCfigure}

The resulting model is summarized \review{schematically} in
Fig.~\ref{fig:compartments}, \review{while a detailed description} of
the model and all parameters \review{are found} in the SI. With
increased model complexity follows explanatory power at the expense of
poorer model identifiability. With this in mind, we exclude all model
refinements that are either missing or are unreliably or incompletely
reported in the data, e.g., age and gender as well as certain refined
states of minor symptoms. We recognize the infection's varying effects
on different age groups \cite{fhm2021delrapportX} and so must accept
that our results remain in an age-averaged regime.

Increased model complexity which, however, \emph{is} required involves
using a mix of dynamic and static parameters, since this allows the
model to respond to functional changes such as societal interventions
\cite{haug2020ranking}, vaccinations and virus mutations
\cite{liu2021reproductive}. We thus let $\beta_t$, \review{that is,}
the infection rate \review{which is} related to the reproduction
number $R_t$, \review{as well as} the infection fatality rate (IFR)
\review{both} be time-varying \review{parameters}. All in all the
problem is then to determine the posterior distribution for 10 static
and two dynamic parameters. The latter are assumed constant for
periods of four weeks but are re-sampled independently for each such
period.

\begin{figure}[tbhp]
  \centering
  \includegraphics[clip = true,
  trim = 0.25cm 8.2cm 0.25cm 8.0cm, 
  width=\figwidth]
  {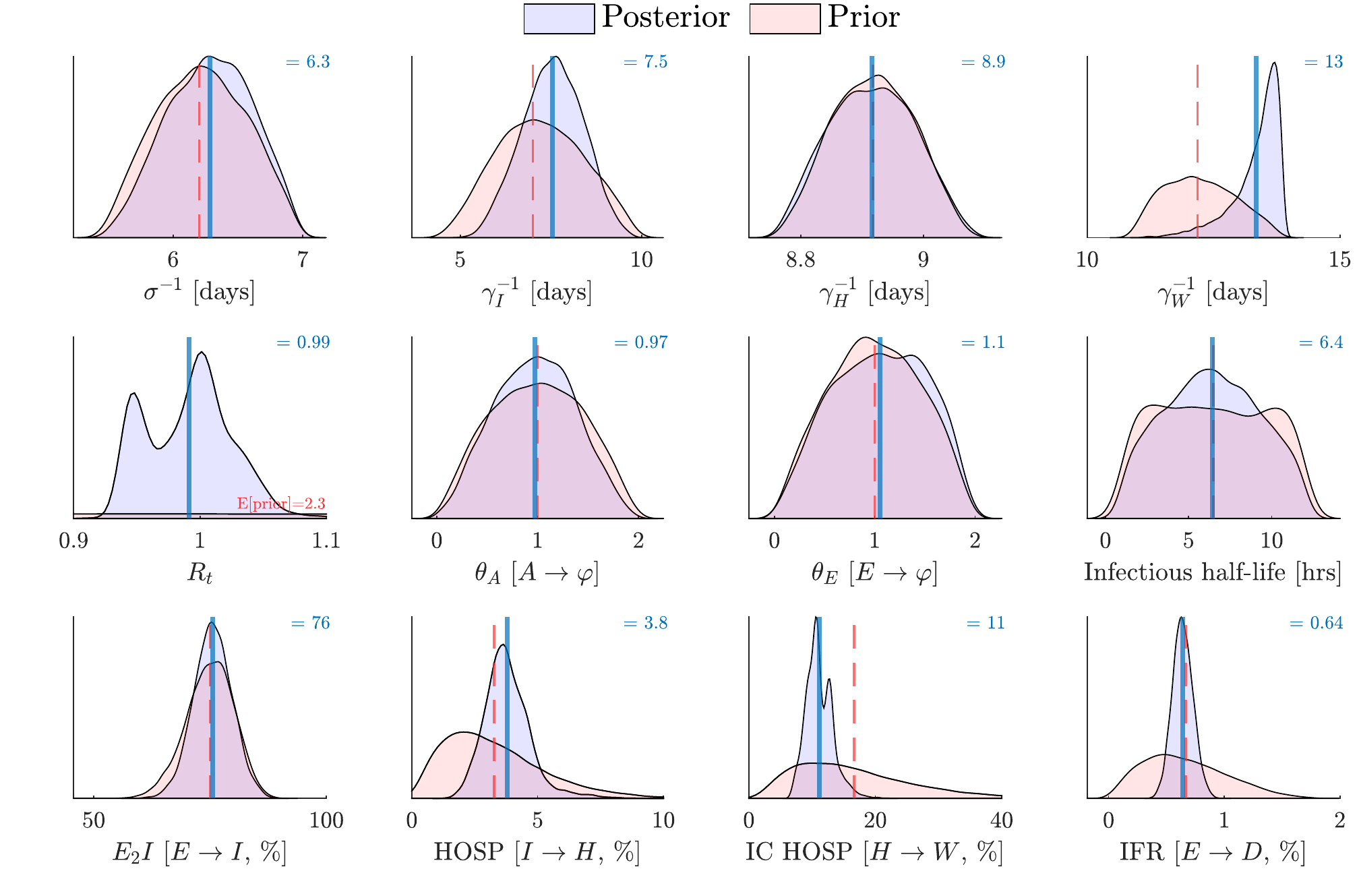}
  \caption{Marginal priors (red) for the model parameters and the
    associated posterior distributions (blue) for the Swedish
    aggregate inferred from publicly available regional data from
    April 1, 2020 to May 30, 2021. The dashed lines indicate the prior
    means, and the full lines and numeric values the posterior
    means. \review{\textit{Top row:} latent period rate $\sigma$, and
      exit rates $\gamma_X$ out of compartment $X \in
      \{I,H,W\}$. \textit{Middle row:} reproduction number and
      contribution to the infectious pressure $\varphi$ from
      compartments $A$ and $E$, respectively, as well as its decay
      (half-life). \textit{Bottom row:} fractions proceeding between
      the indicated compartments.} The reproduction number $R_t$ and
    the IFR are dynamic parameters and a temporal average is displayed
    here.}
  \label{fig:posterior_sweden}
\end{figure}

Clearly, well-chosen priors are required to make the problem definite.
Considerable work went into constructing and continuously updating our
priors using published research and public registries; the final
priors are displayed in Fig.~\ref{fig:posterior_sweden}, with a
complete list of priors and prior predictive estimates given in the
SI.

\subsection*{Approximate Likelihoods through Linear filtering}



We understand the compartment model as a continuous-time Markov chain
(CTMC) over an integer lattice counting the number of individuals in
the different compartments. Hence the waiting time for exiting one
compartment is exponentially distributed of mean rate $1/\lambda$, and
the number of individuals which exit in a small window of time
$[t,t+\Delta t)$ is Poissonian $\sim \text{Po}(n_t \lambda \Delta t)$,
with $n_t$ the number of individuals at time $t$ in the
compartment. If the number of individuals is large enough, this
transition can be approximated by the normal distribution
$\sim \mathcal{N}(n_t \lambda \Delta t,n_t \lambda \Delta t)$, which
can be directly translated into a contribution to the state update
matrices $(F_k,Q_k)$ of a discrete-time Kalman filter with the
equations of state
\begin{align}
  x_{k+1} &= F_k x_k + w_k, \qquad w_k \sim \mathcal{N}(0,Q_k),
\end{align}
with $k$ the discrete time index \review{corresponding to days}.

The linear filter allows for an approximation of the intractable
likelihood of a parameter proposal $\tilde{\Theta}$, namely the
marginal filter likelihood,
\begin{align}
  p_{\tilde{\Theta}} (y) &= \prod_{k=0}^T \mathcal{N}(y_k| H_k \hat{x}_{k|k-1}, Q_k),
\end{align}
\reviewb{where $y_k$ is the data at time $k$, $H_k$ the observation
  matrix which maps the state to an observation, and where
  $\hat{x}_{k|k-1}$ is the Kalman state predictor at time $k$ given
  data until time $k-1$.}  In our application we rely on the
likelihood to produce an approximate sample $(\Theta_i)$ from the
posterior using the Adaptive Metropolis (AM) algorithm
\cite{haario2001adaptive}. The role of the marginal likelihood is
remindful of a \emph{synthetic likelihood} \cite{wood2010statistical}
and we refer to the specific combination of a Kalman marginal
Likelihood and Adaptive Metropolis as KLAM.

\subsection*{\reviewb{Improved transmission rate estimation}}



The achievable temporal resolution of the Bayesian parameter estimates
provided by Metropolis sampling is limited by both computational
complexity and the amount of data. The procedure described so far
yields static reproduction number estimates for each four-week period
and with comparably large spread. To obtain more fine-grained
estimates, a different approach is needed. Since the reproduction
number is the dominating parameter of the dynamics, a more highly
resolved estimate of the reproduction number is particularly
useful. In terms of the parameters of the model, this corresponds to
improving the estimation of $\beta_t$. Therefore, a daily estimate
$\beta(t_k) = \beta_k$ is calculated using dynamic optimization
techniques.

Dynamic optimization has been used for estimating the reproduction
number of the COVID-19 outbreak also by others \cite{RC20}. When
combined with an existing posterior distribution, care must be taken
to avoid overconfidence from using the same data twice. For this
reason, we do not attempt to derive an improved posterior distribution
of $\beta_t$, but instead a single marginal time-dependent maximum
likelihood estimate is sought, where the rest of the posterior
distribution is ``frozen'' \review{and consequently the parameter
  uncertainty is the same in absolute terms}. The same logarithmic
marginal likelihood that is used in the Kalman filter is utilized for
this purpose, now understood as a quadratic cost function. Here,
however, the deviation between measurements and the outputs from the
mean-field dynamics is minimized, i.e., a shortened formulation
compared to the filter is employed since the Kalman correction step is
neglected. To avoid fast variations in $\beta_k$, a regularizing term
penalizing square gradients is also added. The resulting optimization
problem can be solved using standard techniques. Further details on
the formulation of the optimization problem and its solution are
provided in the SI.

\review{Apart from providing more detailed information, this procedure
  yields improved confidence in the Bayesian workflow since it
  supports synthetic data with a known truth to be simulated in an
  off-line fashion. Our Bayesian inversion may thus be employed a
  second time in order to estimate bias or sensitivities for various
  estimates of interest, next to be described.}

\subsection*{\reviewb{Assessing the quality of the approximate posterior}}



In real applications, a ``true'' or a ``best'' parameter posterior
$\Pr^*$ is usually unknown. Evaluating the stability and the quality
of the approximate posterior $\tilde{\Pr}$ is unfortunately often
overlooked. We suggest employing a \emph{parametric bootstrap}
approach as in \cite{engblom2020bayesian} to assess the error between
samples from the true and the approximated posterior
$\tilde{E} := \tilde{\Theta}-\Theta^*$, where $\Theta^* \sim \Pr^*$
and $\tilde{\Theta} \sim \tilde{\Pr}$. Denote by
$\theta^* := \Expect[\Theta^*]$, the minimum mean square error
estimator (MMSE) of an assumed truth. Decomposing the mean square
error around this value we find
\begin{align}
  \label{eq:MSEdecompose}
  \tilde{e}^2 &:= \Expect[(\tilde{\Theta}-\theta^*)^2] =
                \underbrace{\Expect[(\tilde{\Theta}-\tilde{\theta})^2]}_{\text{Variance}}
                +\underbrace{(\tilde{\theta}-\theta^*)^2}_{\text{Square bias}\, =: \,\tilde{b}^2},
\end{align}
where $\tilde{\theta} := \Expect[\tilde{\Theta}]$ is the MMSE of
$\tilde{\Theta}$.

Formally, this still requires samples from the true posterior when
estimating the bias. We approximate this via a bootstrapped estimator
using a sample of $N_{\text{boot}}$ synthetic data sets generated from
the MMSE of the approximate posterior. The generative simulator
requires daily estimates of $(\beta_k)$ or else the synthetic data
quickly drifts off compared to the observations, and thus this
technique ultimately hinges on the highly resolved marginal estimator
described above. Posterior samples may then be generated for each
synthetic set, yielding now \emph{a set} of samples
$\hat{\Theta}_i \sim \hat{\Pr}_i$, which allows for the use of the now
tractable bias estimator
$\hat{b}_i := \Expect[\hat{\Theta}_i]-\Expect[\tilde{\Theta}]$. Our
final estimator is then an average over these synthetic sets;
$\tilde{b}^2 \approx N_{\text{boot}}^{-1} \sum_i \hat{b}_i^2$ in
\eqref{eq:MSEdecompose}. While up to $\numprint{196000}$ samples from
each posterior were used to compute point estimators and credible
intervals (CrIs), bootstrap replicas are much more costly to process
so we used $N_{\text{boot}} = 3$, and mainly relied on the bias
estimator to diagnose non-robustness in point estimators. That is, a
point estimator with CrI $A$ of order $\alpha = 68\%$, say, and with
bias estimate $\tilde{b}$ is considered less robust whenever
\begin{align}
  \label{eq:ptoverlap}
  \tilde{b} \geq 0.5\diam(A).
\end{align}

The bootstrap posterior densities can be used in aggregate form for
bias estimation and check of estimator robustness as just described,
or for a related check of credible interval robustness. Consider two
CrI intervals $A$ and $B$ of order $\alpha$, e.g., $\alpha = 68\%$,
and where $B$ is a bootstrap replicate of $A$. A basic measure of the
robustness of $A$ is the level of overlap between $A$ and $B$ and one
can reasonably require the same overlap as the indicated order
$\alpha$, i.e., to require
\begin{align}
  \label{eq:overlap}
  \diam(A \cap B) \ge \alpha \diam(A \cup B).
\end{align}
If this criterion is satisfied, then a random variable drawn uniformly
from $A \cup B$ has probability $\geq \alpha$ to also be in
$A \cap B$. The robustness checks
Eqs.~(\ref{eq:ptoverlap})--(\ref{eq:overlap}) are explicitly reported
in Tab.~\ref{tab:IFR}, but were also routinely employed when
evaluating various results.


\section*{Results}



Our model was used for weekly reported predictions within CRUSH Covid.
Prior to each report we carried out a model updating procedure: new
data were pulled from public repositories and screened for
contradictory or incorrect values. The posterior was sampled by KLAM
using an initialization either from an estimated initial state as
described in the SI and developed for this very purpose, or simply
using stored state samples. The latter allows for faster sampling as
it reduces the burn-in period: about 24 hours of compute for 21
regions and a year's worth of data on a 4-core laptop was then reduced
to a few hours. The final posterior model was queried for
one-week-ahead forecasts with uncertainty bounds for the $(H, W, D)$
triple. We also continuously evaluated the previous week’s predictions
against the up-to-date data in the same round.

\subsection*{Posterior prediction}



In Fig.~\ref{fig:posterior_sweden}, we display the prior distributions
together with the resulting Swedish \emph{aggregate} posterior, i.e.,
the population weighted average of the individual posterior of each of
Sweden's 21 regions. Several priors are clearly very similar to their
respective posteriors, e.g., the latent period rate $\sigma$ and the
symptomatic period rate $\gamma_I$. This is expected and simply
indicates little information in the observations for these parameters
relative to the prior. Our data also cannot improve on the prior for
the share of spread from exposed (pre-symptomatic) and asymptomatic
individuals (parameters $\theta_E$ and $\theta_A$, respectively). This
is simply due to the fact that we have no continuously reported data
neither for $E$ nor for $A$. On the same note we had to rely on a
directed study \cite{byrne2020inferred} to define the prior for the
parameter $E_2I$, i.e., the fraction of exposed individuals who
eventually develop symptoms.

Of more interest are the parameters that govern the fraction that
transition to a worsened state of the disease: $I \rightarrow H$ and
$H \rightarrow W$. Our results show that $3.8\%\, [2.2,5.9]$ (95\%
Credible Interval (CrI)\footnote[7]{We indicate Credible Intervals (CrI)
  by square brackets $[\cdot,\cdot]$ and Confidence Intervals (CI) by
  regular parentheses $(\cdot,\cdot)$, unless otherwise specified.})
of the symptomatic individuals require hospital care, and
$11.2\%\, [7.6,16.1]$ of the hospitalized patients require intensive
care. During the considered period, the Swedish Public Health Agency
(PHA) published five point estimates of those same fractions. The
relevant demographic average of these are $I \rightarrow H$:
$3.0\%\, (2.4, 3.6)$ and $H \rightarrow W$: $14.3\%\, (9.8, 18.8)$
(mean $\pm\, 2$ std, $n = 5$) respectively
\cite{fhm2021delrapportX}. Additionally, \cite{salje2020estimating}
found similar estimates in France $I \rightarrow H$:
$2.9\%\, [1.7, 4.8]$ and $H \rightarrow W$: $19.0\%\, [18.7,
19.4]$. We could have used the earlier of those point estimates to
improve on the corresponding priors, however, as validation
possibilities are scarce, we decided to rather use them for this
purpose instead.

\begin{figure}
  \centering
  \includegraphics[clip = true,
  trim = 4.5cm 9.3cm 4.5cm 9.5cm,
  width=\figwidth]{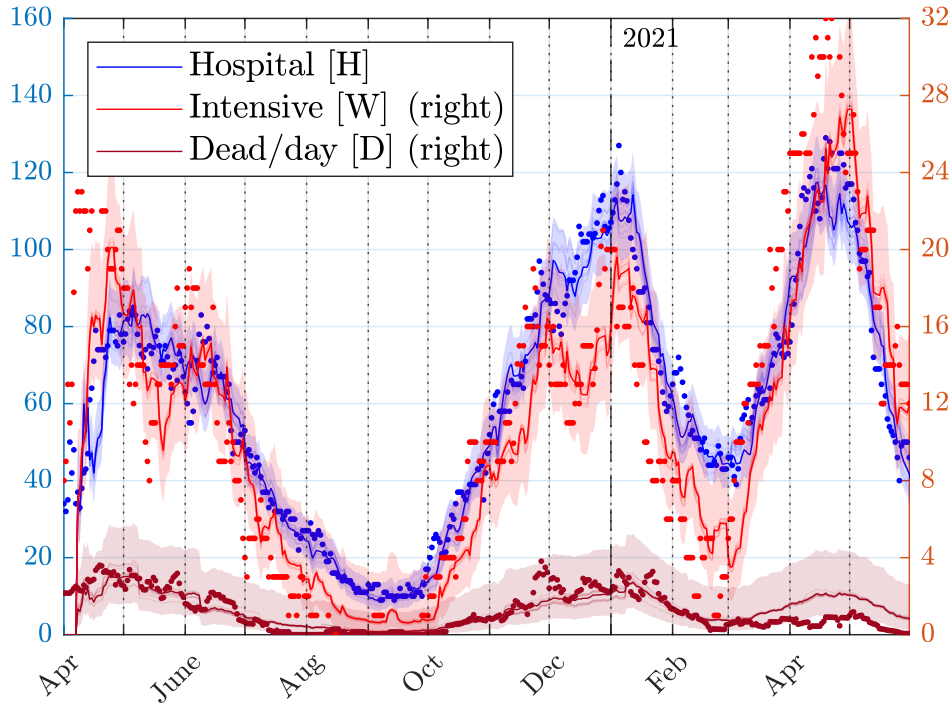}
  \caption{Seven-day ahead prediction in the Uppsala region. Shaded
    area shows 68\% CrI, and the points are the observations. The
    dotted vertical lines indicate the four-week periods used for the
    dynamic rates.}
  \label{fig:lag_uppsala}
\end{figure}

The regional models were used for posterior predictions on a weekly
basis, e.g., one 7-day ahead prediction \review{for the Uppsala and
  Stockholm regions, respectively,} and one \review{in aggregate} for
the entire nation. In Fig.~\ref{fig:lag_uppsala} this is exemplified
over a longer period together with the actual outcome. The performance
of the weekly predictions which were reported live ($N = 25$) is
presented in Tab.~\ref{tab:PastWeekly}. Each prediction included a
mean with 68/95\% CrI. The week after publishing, we evaluated the
prediction against the then-available data. The predictions performed
better in the medium sized region Uppsala and notably, the predictions
for casualties were poor in the larger region Stockholm. \review{The
  observed misfit in casualties eventually lead us to reconsider the
  role of the IFR parameter, initially just a static parameter, and we
  decided to let it be dynamic as described previously. In the live
  reporting in Tab.~\ref{tab:PastWeekly}, only 9 of the 25 reports
  allowed for a dynamic IFR.} Another possible reason for the poorer
prediction performance \review{in the Stockholm region} could be that,
since \review{this} region contains three large hospitals, the greater
heterogeneity in terms of reporting makes identification more
challenging. One can rightly question if smaller sub-regions should
rather be modeled here, but we did not have access to the data to
drive such a model.

\reviewb{In the SI we further compare the quality of the Kalman
  predictor with that of a simpler regression-based estimator. The
  latter provides accurate mean-square predictions and is very fast to
  evaluate. The overall advantage with our approach lies rather in
  that the Bayesian posterior model itself can be investigated for
  further epidemiological insight, next to be discussed.}

\begin{table}[htp]
  \centering
  \begin{tabular}{lrrr}
    \hline
    & Hospital (H) & Intensive (W) & Death (D) 	\\
    \hline
    Uppsala (68\%) &  76 &  72 &  68 	\\
    Stockholm &  64 &  64 &  36 	\\
    Sweden &  88 &  44 &  32 	\\
    \hline
    Uppsala (95\%) & 100 & 100 &  96 	\\
    Stockholm &  84 &  92 &  68 	\\
    Sweden &  96 &  92 &  68 	\\
    \hline
  \end{tabular}
  \caption{Frequency (\%) of all weekly reported predictions that fell
    inside of (68/95\% CrI, 7 days ahead), evaluated on the following
    week (N~=~25). Note that the Uppsala results are very close to the
    ideal 68/95\% outcome.}
  \label{tab:PastWeekly}
\end{table}

\subsection*{Posterior hidden state estimation}

\begin{figure}
  \centering
  \includegraphics[clip = true,trim = 4.5cm 9.3cm 4.5cm 9.5cm,
  width=\figwidth]{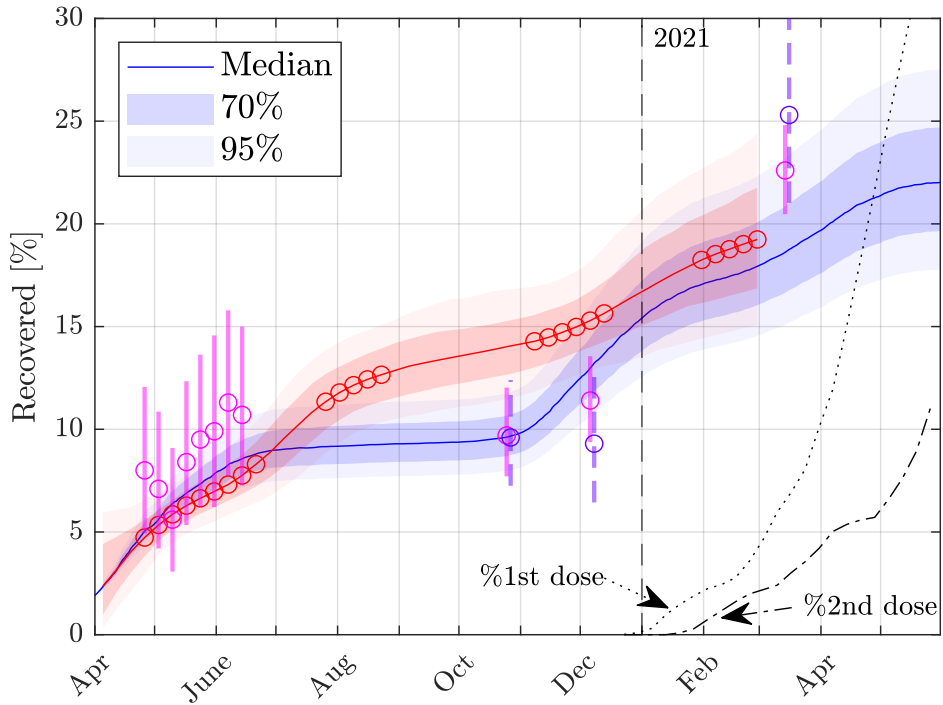}
  \caption{The fraction of recovered individuals in the Stockholm
    region. Our model (blue), is set to 2.35\% on April 5, 2020
    (matching \cite{castro2021seropositivity}). Serology-based
    Bayesian model predictions \cite{castro2021seropositivity} (red)
    which reported 70/95\% CrIs. Estimated mean prevalence of
    antibodies found in spared blood samples from outpatient care,
    with 95\% CI (solid, pink) and blood donors (dashed, purple)
    \cite{fhm2021pavisning}. On 2020, December 27, the Swedish
    vaccination campaign started and both the number of delivered
    first doses (dotted) and second doses (dash-dot) are indicated.}
  \label{fig:recovered_stockholm}
\end{figure}

The posterior model can also be used as a kind of ``Bayesian twin''
and estimate quantities that are otherwise very difficult to
approach. For example, we can readily estimate the number of
individuals that have contracted the disease and survived; these
individuals are the ones that could potentially have developed
antibodies that are detectable in serology tests. In
Fig.~\ref{fig:recovered_stockholm}, we visualize several reported
results for Stockholm \cite{castro2021seropositivity,
  fhm2021pavisning} and our estimates. Those estimates compare very
well, and notably so given that no screening data was used by our
method.

\begin{figure}[!htpb]
  \begin{subfigure}{0.49\textwidth} \includegraphics[clip = true,
    trim = 4.5cm 9.3cm 4.2cm 9.5cm,
    width=1\textwidth]{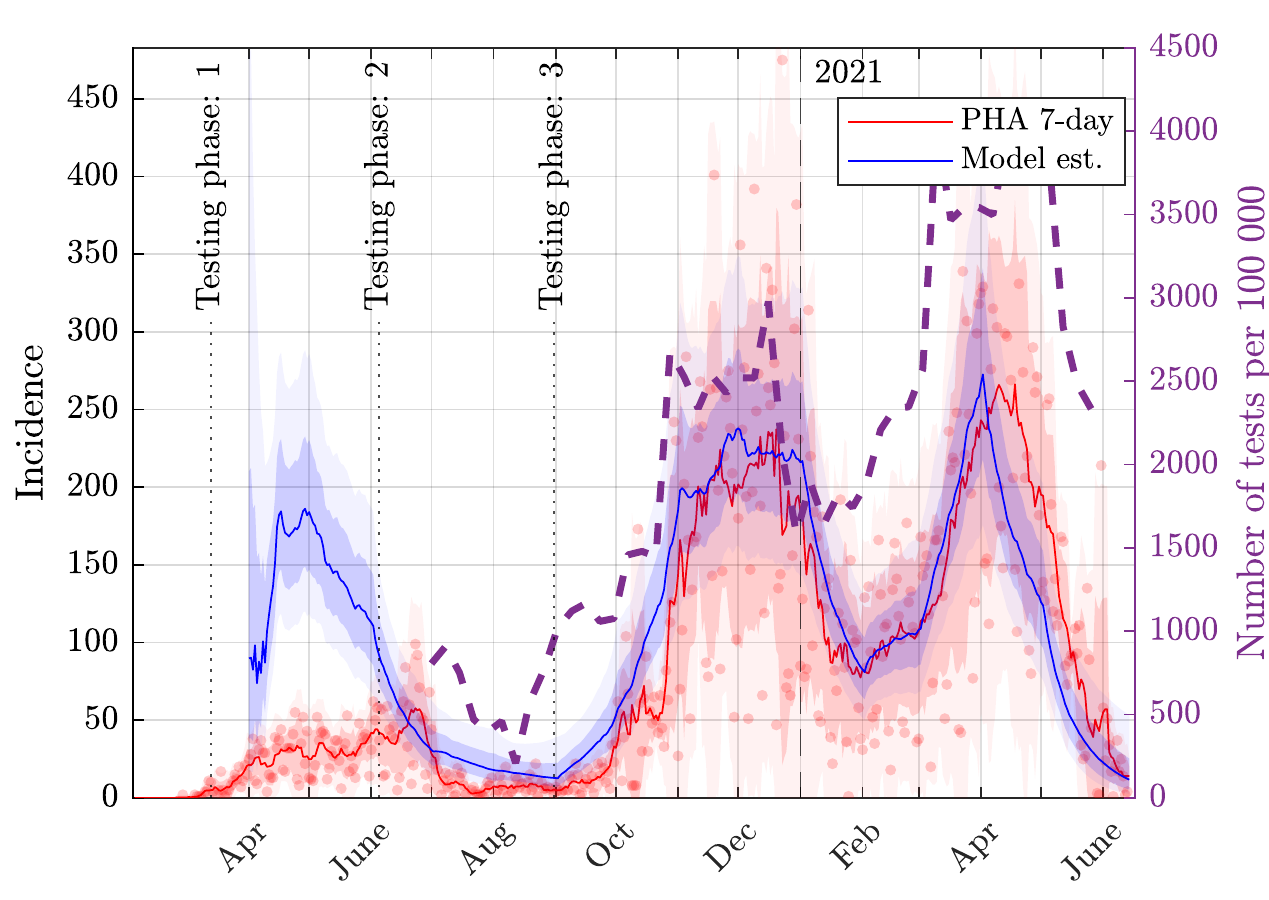}
  \end{subfigure} \hfill
  \begin{subfigure}{0.49\textwidth} \includegraphics[clip = true,
    trim = 4.5cm 9.3cm 4.2cm 9.5cm,
    width=1\textwidth]{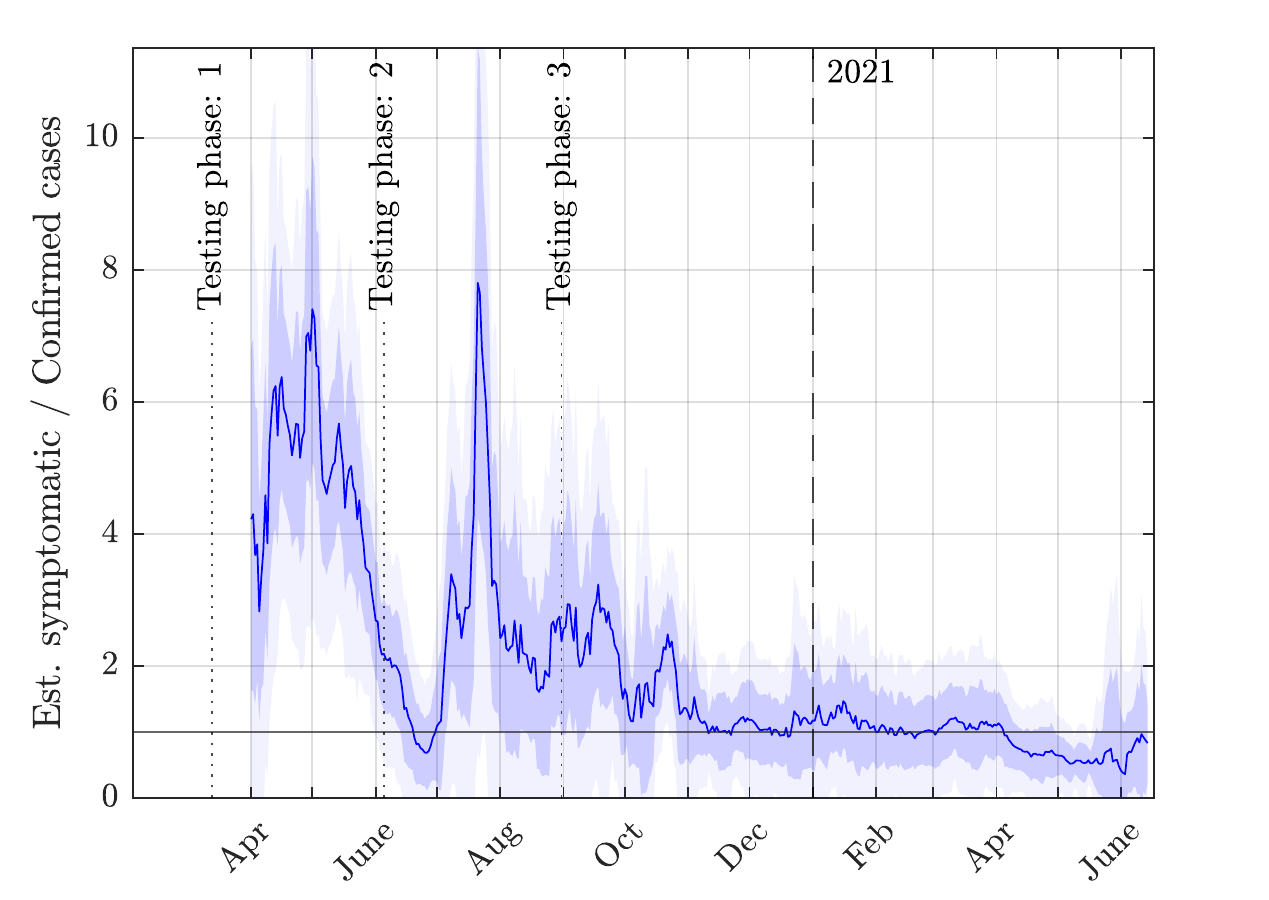}
  \end{subfigure}
  \caption{\textit{Left:} the estimated symptomatic incidence from our
    model (blue with [68,95]\% CrIs shaded) and the confirmed cases by
    PHA (red line 7-day smoothed and uncertainty from rolling
    $\pm [1,2]$ standard deviations shaded, with data as points) for
    Uppsala. The right axis gives the number of tests per
    \numprint{100000} inhabitants per week (dashed). The total number
    of tests administered during Testing phase 3 was approximately
    \numprint{360 000}. The dotted vertical lines \review{mark the
      boundaries of testing phases as defined by policy changes
      affecting the testing volumes}. \textit{Right:} the ratio
    between the model's symptomatic incidence and the confirmed cases
    incidence indicates the proportion of cases that are missed by
    testing. Values above one thus hint at an underreporting of
    symptomatic cases.}
  \label{fig:Iinc_uppsala}
\end{figure}

The posterior model can also estimate the symptomatic incidence in the
same vein. Fig.~\ref{fig:Iinc_uppsala} illustrates our estimated
symptomatic incidence and the reported number of positive RT-PCR tests
by the PHA for Uppsala. As testing increases, the ratio between our
estimate and the positive tests oscillates around one, indicating that
the testing at that time captures most of the symptomatic infected.

These two examples demonstrate that our computational model can be
used to effectively estimate hidden variables at the same regional
resolution as supported by data. The financial cost for this kind of
monitoring would of course be a tiny fraction compared to any
alternatives based on testing.

\subsection*{Marginal risks}


Recall that the IFR is defined as the proportion of deaths among
infected individuals, i.e., including asymptotic cases. By design, our
model relies on an IFR which is constant over four weeks. Our
estimated IFR for Sweden stayed relatively constant over the period
May 2020--November 2020 at $0.69\%\, [0.11, 1.5]$ (95\% CrI). For
comparison, in April 2020, the PHA published an early estimate of
$0.58\%\, (0.37, 1.05)$ (95\% CI) for Stockholm, Sweden
\cite{fhm2020IFR}. However, this estimate relies on initial
assumptions on the number of undetected cases that seem unjustified
when compared to later findings \cite{rippinger2021evaluation,
  irons2021estimating}. It also assumes the relatively younger
Stockholm demographics rather than the national one, and is therefore
an underestimate of the national IFR. A later Sweden-wide IFR estimate
of $0.76\%\,[0.65,0.87]$ was published in November 2020
\cite{garcia2021temporal} and aligns well with our estimate
above. Although our 95\% CrI is comparatively wide, this is partially
due to modeling each period independently and without any
regularization in the transition between periods.

Tab.~\ref{tab:IFR} summarizes a few bi-monthly estimates for the
period after October 2020. The first two entries in the table overlap
the estimates from \cite{covid2020variation}: Nov 2020
$= [0.60, 1.46]\%$, and Jan 2021 $= [0.56, 1.44]\%$. Clearly, the IFR
was trending downwards and this is also known to be the case in
Stockholm \cite{garcia2021temporal} as well as for the world in
general \cite{covid2020variation}.

\begin{table}[htp]
  \centering
  \begin{tabular}{l r r r }
    \hline
    & Stockholm& Uppsala& Sweden \\\hline
    Oct+Nov& \footnotemark[3]\CI{0.12}{0.34}{0.66}{0.81}{0.91}& \CI{0.08}{0.24}{0.49}{0.74}{0.96}& \CI{0.10}{0.27}{0.55}{0.88}{1.30} \\
    Dec+Jan& \footnotemark[3]\CI{0.61}{0.71}{0.84}{1.00}{1.10}& \footnotemark[3]\CI{0.10}{0.28}{0.59}{0.92}{1.20}& \CI{0.20}{0.55}{0.88}{1.30}{1.60} \\
    Feb+Mar& \footnotemark[4]\footnotemark[3]\CI{0.07}{0.18}{0.35}{0.50}{0.61}& \CI{0.08}{0.23}{0.47}{0.70}{0.89}& \CI{0.08}{0.22}{0.44}{0.81}{1.20} \\
    Apr+May& \CI{0.07}{0.19}{0.35}{0.49}{0.59}& \CI{0.08}{0.22}{0.45}{0.67}{0.86}& \CI{0.07}{0.18}{0.35}{0.56}{0.92} \\
    \hline
  \end{tabular}
  \begin{tabular}{l}
    {\scriptsize \footnotemark[4] The estimated bias is large compared
    to the 68\% CrI (see \matmet).} \\
    {\scriptsize \footnotemark[3] The 68\% CrIs of the posterior and the bootstrap
    replicate do not share at least a 68\% overlap (see \matmet).}
  \end{tabular}
  \caption{Bi-monthly estimated IFR [\%] with 68\% CrI from October 2020
    to May 2021. }
  \label{tab:IFR}
\end{table}

Of interest is also the \textit{case fatality risk} (CFR)
\cite{kelly2013case}, i.e., the risk of death conditioned on being
diagnosed with the disease. We more generally define CFR$_X$ as the
proportion of deaths expected given a certain number of individuals in
compartment $X \in \{I, H, W\}$. Note that CFR$_I$ involves the number
of symptomatic individuals which formally is not the same thing as the
number of cases confirmed by testing.

From our posterior we may directly estimate the national average CFRs
to be
$\{
\text{\CIedge{0.72}{0.86}{0.99}{1.1}{1.3}},
\text{\CIedge{16}{16}{17}{18}{19}},
\text{\CIedge{34}{34}{35}{35}{36}}
\}\%$ (95\% CrIs) for CFR$_I$,
CFR$_H$, and CFR$_W$.  By comparison, \cite{alimohamadi2021case}
offers the estimates CFR$_I$ $= (1.0, 3.0)\%$, CFR$_H$
$= (9.0, 17.0)\%$, and CFR$_W$ $= (24.0, 51.0)\%$ (95\% CI).

Another way to investigate these risks is by running the posterior
filter across our data to produce an estimate for the number of
deceased per compartment. Until March 23, 2021 we find for all of
Sweden that
$\{D_I,D_H,D_W\} =\{ \text{\CIedge{1786}{3502}{5219}{6935}{8652}},
\text{\CIedge{3612}{4733}{5854}{6975}{8096}},
\text{\CIedge{1126}{1751}{2376}{3002}{3627}} \}$ (95\% CrIs). From
NBHW data we may estimate those same numbers to be
$\{\numprint{8335}, \numprint{4102}, \numprint{935}\}$
\cite{ss2020avlid}, after a scaling of about 5\% to arrive at the same
total number of dead as in our dataset ($D =
\numprint{13372}$). Apparently, our model overestimates the deaths
under ICU and to some extent also at hospital, while our estimate for
deaths outside hospital is \review{underestimated} compared to this
data point. This could likely improve given more detailed data
sources, including, e.g., improved records of COVID-related deaths and
hospital care outcomes.

\subsection*{Reproduction number estimates}

The \emph{reproduction number} provides an essential insight into the
future development of the spread of a disease in a population. An
accurate estimate of this number is vital to support rational public
health decision-making and to inform the general public. Temporal
variations in the reproduction number are caused by socio-behavioral,
environmental, and virological and biological factors
\cite{delamater2019complexity}. The dynamics of the reproduction
number is therefore significantly faster than those of most other
parameters, which was also the motivation behind our development of an
improved marginal estimator via dynamic optimization techniques as
described previously.

\begin{figure}
  \centering
  \includegraphics[clip = true,
  trim = 2.8cm 9.0cm 2.8cm 9.0cm,
  width=\figwidth]{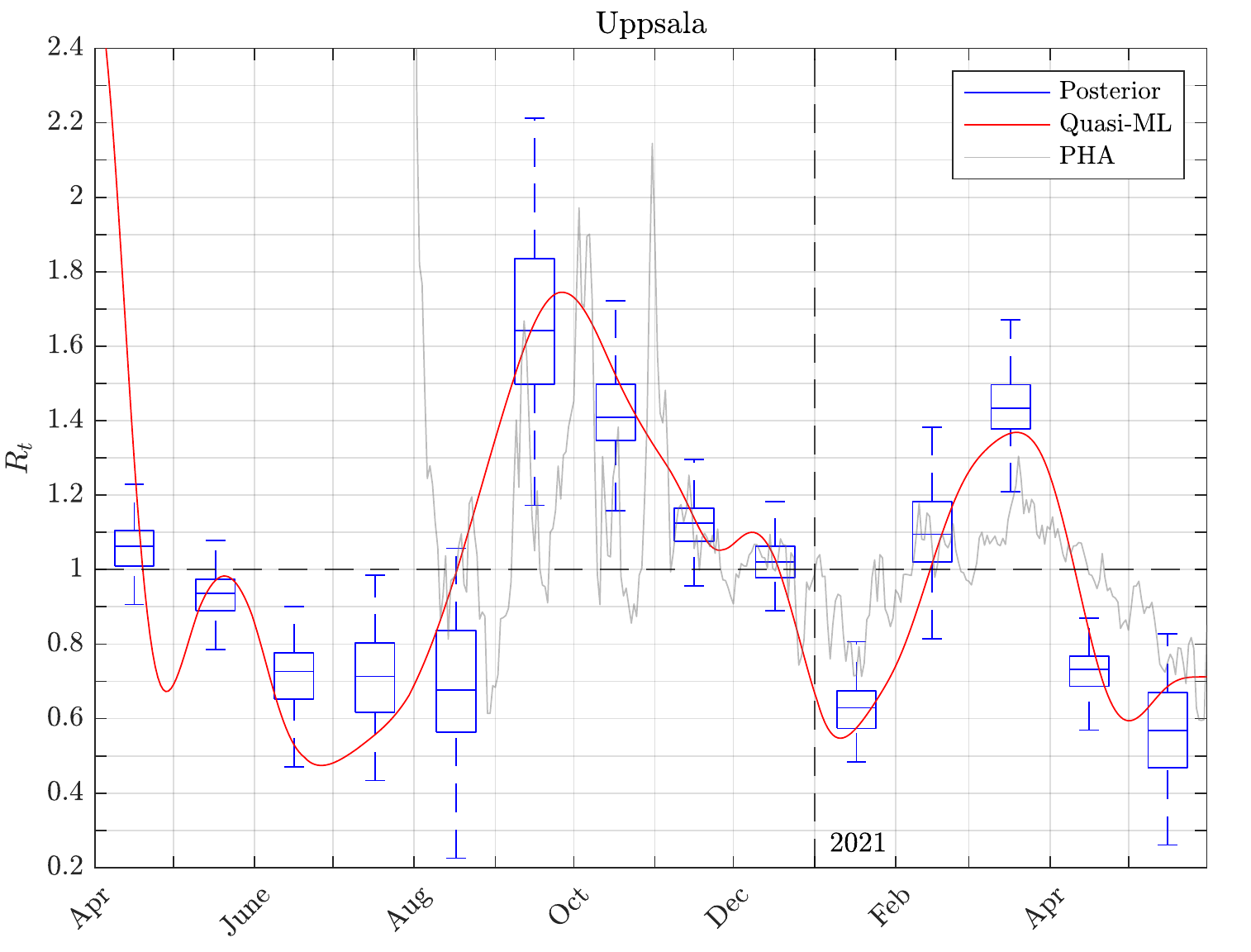}
  \caption{Reproduction number estimates for the Uppsala region. The
    Bayesian posterior yields monthly estimates (box-plot) while the
    quasi-ML estimator is daily (solid). For comparison also the PHA's
    estimates based on screening from of a total of \numprint{367200}
    RT-PCR tests during the period 5 August, 2020--May 26,
    2021. \review{Note that reproduction numbers are model specific
      and the comparison here is therefore mainly qualitative.}}
  \label{fig:Rposterior}
\end{figure}

Reproduction number estimates have been calculated in this way for the
whole duration of the parameterized period for all regions in
Sweden. In Fig.~\ref{fig:Rposterior}, we present our results for
Uppsala together with the testing-based estimate produced by the PHA
for the same period. The cost-effectiveness of our estimate is
apparent here since the results are similar, but the PHA estimate
relies on costly incidence data.

\section*{Discussion}



The pathogen SARS-CoV-19 and the subsequent pandemic resulted in an
explosion in research related to COVID-19: research on data collection
and interpretation, modeling and forecasting, as well as scenario
generation. From the beginning of the outbreak, data have been
instrumental in understanding the disease dynamics \cite{li2020early,
  lavezzo2020suppression, wu2020nowcasting}. With increasing amounts
of cases and recorded patient data, statistical models for diagnosis
and prognosis were quickly developed
\cite{wynants2020prediction}. Combining in-patient data with other
covariates, the use of digital technologies for disease surveillance
is now possible \cite{abad2021digital}, notably with an impact also
for COVID-19 \cite{menni2020real, drew2020rapid, rossman2020framework,
  kennedy2022app}. Challenges connected to the collection and
distribution of available data where met with specialized tools for
decentralized publishing \cite{guidotti2020covid, reinhart2021open}
and anonymized mobility data were also in active use
\cite{imran2022tbcov, ilin2021public}.

We have devised a detailed Bayesian model for the regional dynamics of
the COVID-19 pandemic in Sweden. A data-centric viewpoint is that,
using \emph{model-based data analysis}, we have gained a thorough
insight into the progression of the COVID-19 pandemic in Sweden in
general, with extra emphasis on the Uppsala Region. The proposed
compartment-based model combined with the novel use of optimal linear
filters turned out to be \review{an effective} information-theoretic
epidemiological tool. The output quality as obtained in our work
compares very well with official estimates gathered in test-based
programs,
cf.~Figs.~\ref{fig:recovered_stockholm}--\ref{fig:Rposterior}. During
the second and third waves in Sweden, at least 10 million RT-PCR tests
were administered at a standard cost of \numprint{1400} SEK
($\approx \$150$) per test \cite{pcr2021skr}. The expenses for such a
\emph{PCR test for all symptomatic}-strategy quickly grows,
underlining the economical advantages of our approach. Clearly, at the
individual level there are several benefits with testing and the
importance of screening as a means to collect initial statistics for
the disease spread cannot be stressed enough. However, our approach
remains very promising as a supporting tool to continue to monitor the
situation when testing is limited due to risk-cost trade-offs.

The resulting model was further processed to output weekly predictions
for health care demands, and also marginal estimators for important
characteristics of the disease such as infection fatality rates and
reproduction numbers. The latter output increased the confidence into
the overall approach through the generation of synthetic data and
parametric bootstrap techniques. Improved data that would have enabled
a higher model precision include (1) a consistently managed incidence
report from randomized testing (not necessarily high volume), and (2)
a higher temporal resolution of hospitalization and intensive care
risks as well as times for treatment in these respective
categories. These statistics could both be collected at a relatively
small cost but would likely improve the precision considerably.

Without relying on public testing strategies, our model-based approach
provided improved \emph{situation awareness} of the progression of the
pandemic. \reviewb{The developed methodology is of highly general
  character and can therefore be expected to be useful in other
  contexts too.} By virtue of the consistent Bayesian framework,
uncertainties are transparently propagated under clear assumptions,
also in the face of potentially hazardous situations. We argue that
this quality makes the techniques developed herein particularly
promising from a communicative perspective.

Our data streams are high in latency, but are on the other hand fairly
low in noise. Low-latency signals, e.g., public screening,
self-reporting mobile apps, or analysis of sewage water, are instead
often more noisy or otherwise biased. Combining these different kinds
of streams provides for excellent decision support and appears
extremely promising for use in tracking regional epidemics at a
near-daily resolution.


\subsection*{Acknowledgments}

This work was financially supported by the Swedish Innovation Agency
Vinnova, by the Swedish Research Council Formas (S.~Engblom), and by
the Swedish Research Council (H.~Runvik and A.~Medvedev).

\subsection*{Author contributions}

SE conceived the research and RM and SE developed the initial forward
model. Linear filters were designed by SE and HR who also developed
and operated the dynamic optimization approach with inputs from AM. SE
developed priors and data pre-processing and RM adapted the Bayesian
sampling techniques, collected data, performed computations, and
prepared the manuscript joint with SE and with inputs from HR. All
authors took part in revising the manuscript.


\bibliographystyle{abbrvnat}
\newcommand*{\doi}[1]{\href{http://dx.doi.org/#1}{doi: #1}}



\renewcommand*{\citenumfont}[1]{SI#1}
\renewcommand*{\bibnumfmt}[1]{[SI#1]}

\let\oldcite\cite
\renewcommand*\cite[1]{\oldcite{SI-#1}}

\appendix

\newpage \onecolumn

\begin{center}
  \LARGE{\textbf{Supporting Information: \\ Bayesian Monitoring of
      COVID-19 in Sweden}}
\end{center}

This supplement contains further explanations and details of the data,
the model, and the computational methodology, as well as some
additional results mentioned in the main article. The SI is organized
in sections as follows:
\begin{description}
\item[Additional supporting results]
  \begin{itemize}
  \item[] 
  \item \textit{Incidence}: our model's estimate of the incidence of
    symptomatic cases compared to the number of confirmed cases by PHA
    in Stockholm (complementing the Uppsala cases in the main
    article). This also produces an estimate of the proportion of
    undetected symptomatic cases.
  \item \textit{$R_t$-estimators:} for a selection of small and large
    regions.
  \end{itemize}
\item[Data] A summary of the various sources and use of data.

\item[Compartment model] Our extended SEIR-model in detail, including
  an explanation of all model parameters.

\item[Priors] The parameter priors for the model and how they were
  determined.

\item[Kalman filter] The linear filter approximation of the
  continuous-time Markov chain, which in turn is our probabilistic
  understanding of the extended SEIR-model.

\item[Posterior sampling] Details of the Metropolis sampling procedure.

\item[Dynamic optimization solution] The procedure for `bootstrap on
  the margin' to improve the temporal resolution of the reproduction
  number estimator.

\item[Additional evaluations]
  \begin{itemize}
  \item[] 
  \item \textit{Posterior robustness:} A comparison of the 21
    regional posteriors, which thus form a natural bootstrap
    population.
  \item \textit{Bootstrap robustness:} The estimated bias and some
    additional quality statistics for the 21 regional posteriors.
  \item \textit{Baseline predictor:} A comparison of the prediction of
    our Kalman filter and a simpler regression estimator.
  \end{itemize}
\end{description}

\subsection*{Reproducibility}

The code as well as the data required for reproducing the results in
the paper are publicly available and can be downloaded at
\url{github.com/robineriksson/Bayesian-Monitoring-of-COVID-19-in-Sweden}. Refer
to the included file {\tt README.md} for more information.


\section*{Additional supporting results}

\subsection*{Incidence estimates: a comparison between model- and
  test-based results}

\begin{figure}[tbhp] \centering %
  \begin{subfigure}{0.49\textwidth} \includegraphics[clip = true,
    trim = 4.5cm 9.3cm 4.2cm 9.5cm,
    width=1\textwidth]{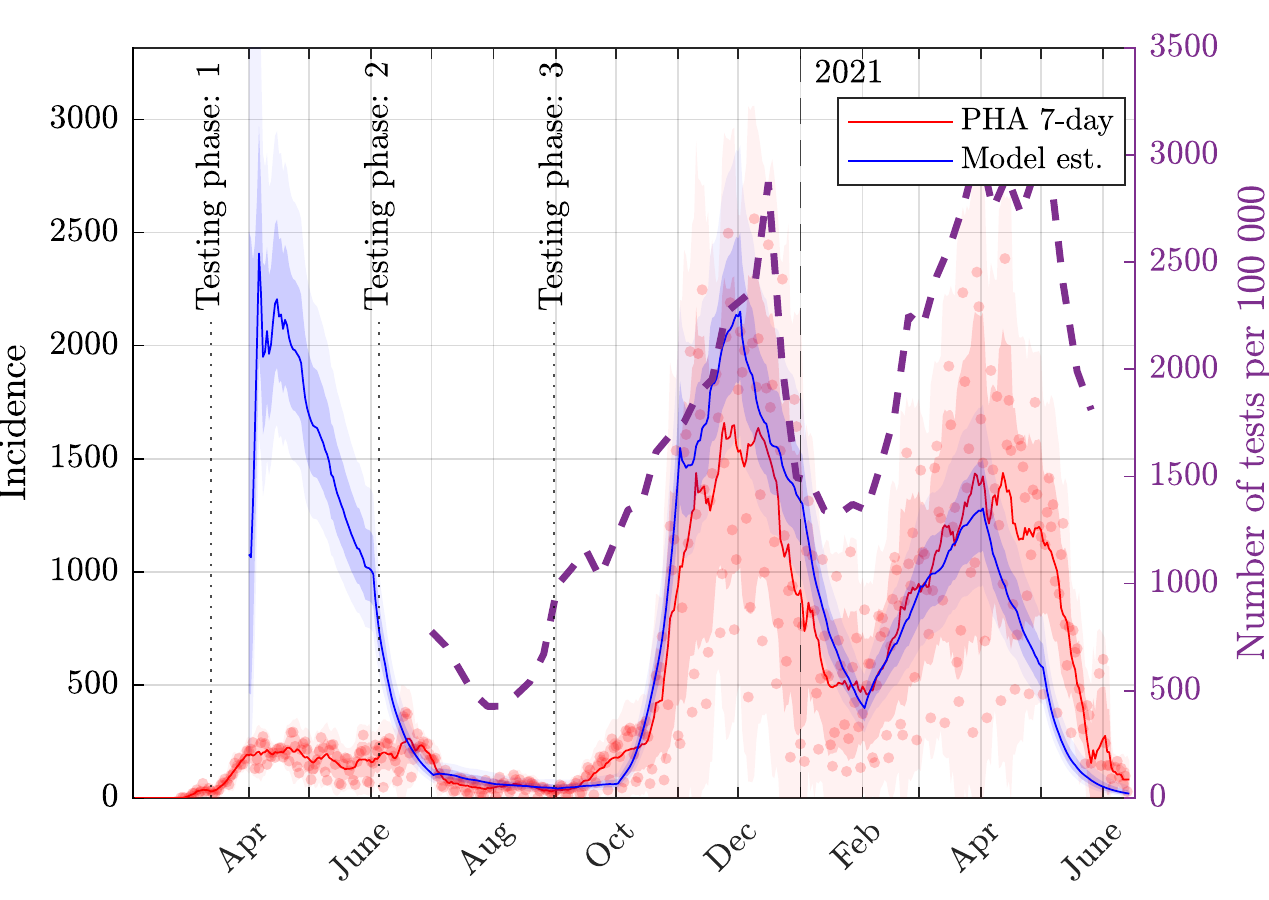}
  \end{subfigure} \hfill
  \begin{subfigure}{0.49\textwidth} \includegraphics[clip = true,
    trim = 4.5cm 9.3cm 4.2cm 9.5cm,
    width=1\textwidth]{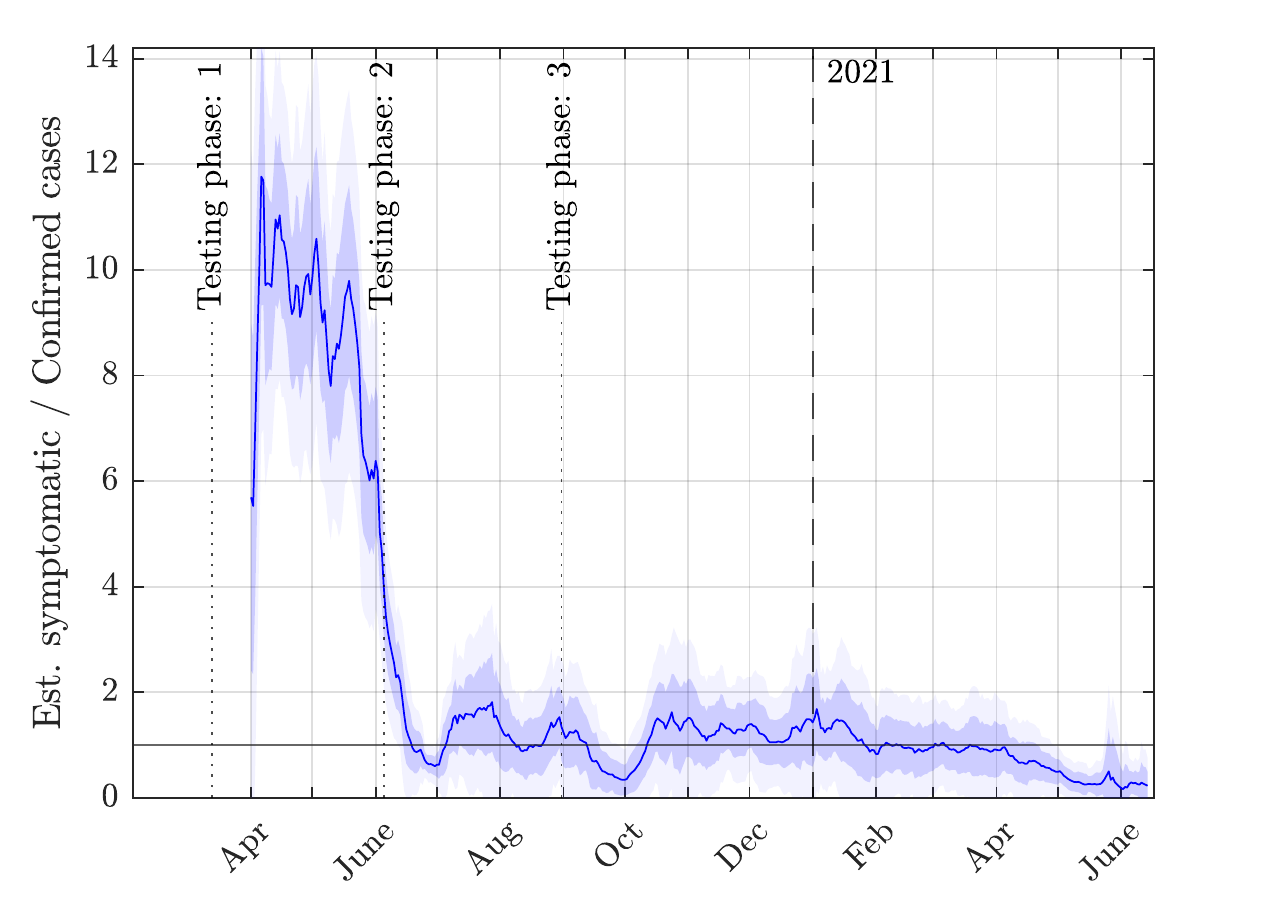}
  \end{subfigure}
  \caption{\textit{Left:} the estimated symptomatic incidence from our
    model (blue with [68,95]\% CrIs shaded) and the confirmed cases by
    PHA (red line 7-day smoothed and uncertainty from rolling
    $\pm [1,2]$ standard deviations shaded, with data as points) for
    Stockholm. The right axis gives the number of tests per
    \numprint{100000} inhabitants per week (dashed). The total number
    of tests administered during Testing phase 3 was approximately
    \numprint{1800000} for Stockholm. The dotted vertical lines mark
    the start of new testing phases that are defined by policy changes
    affecting the testing volumes. \textit{Right:} the ratio between
    the model's symptomatic incidence and the confirmed cases
    incidence indicates the proportion of cases that are missed by
    testing.}
  \label{fig:Iinc}
\end{figure}

In the main paper, we illustrate how the posterior model can estimate
the incidence of symptomatic cases, which is a hidden state and thus
does not correspond directly to data. Here we display the
corresponding estimates for Stockholm, see Fig.~\ref{fig:Iinc}. The
reported tests are smoothed by a 7-day Savitzky-Golay filter and the
estimated uncertainty is the corresponding rolling standard
deviation. After testing phase 3, the ratio between our estimate and
the positive tests oscillates around one, indicating that the testing
at that time captures most of the symptomatic infected. After April
2021, our model estimates a somewhat smaller symptomatic incidence
than what was captured in actual tests. A possible explanation for
this is that at smaller incidence, an increasing proportion of
asymptomatic cases are included in the test pool, e.g., from regular
screening of hospital personnel and similar.

Through this method of exploring the data we can also confirm our
early suspicion that the incidence data signal was poor during Testing
phase 1 and in part during phase 2. The noise in the screening data
explains the nervous behavior of the PHAs reproduction number
estimates as seen in Figs.~\ref{fig:Rposterior} and
\ref{fig:Rposterior_largeNsmall}. At the start of the second wave, and
during Testing phase 3, this signal was much more accurate albeit at
huge economical costs.

\subsection*{$R_t$-estimators}

\begin{figure}
  \begin{subfigure}{.3\textwidth}
    \includegraphics[clip=true, trim = 2.8cm 9.3cm 2.8cm 9.0cm,
    width=1\linewidth] {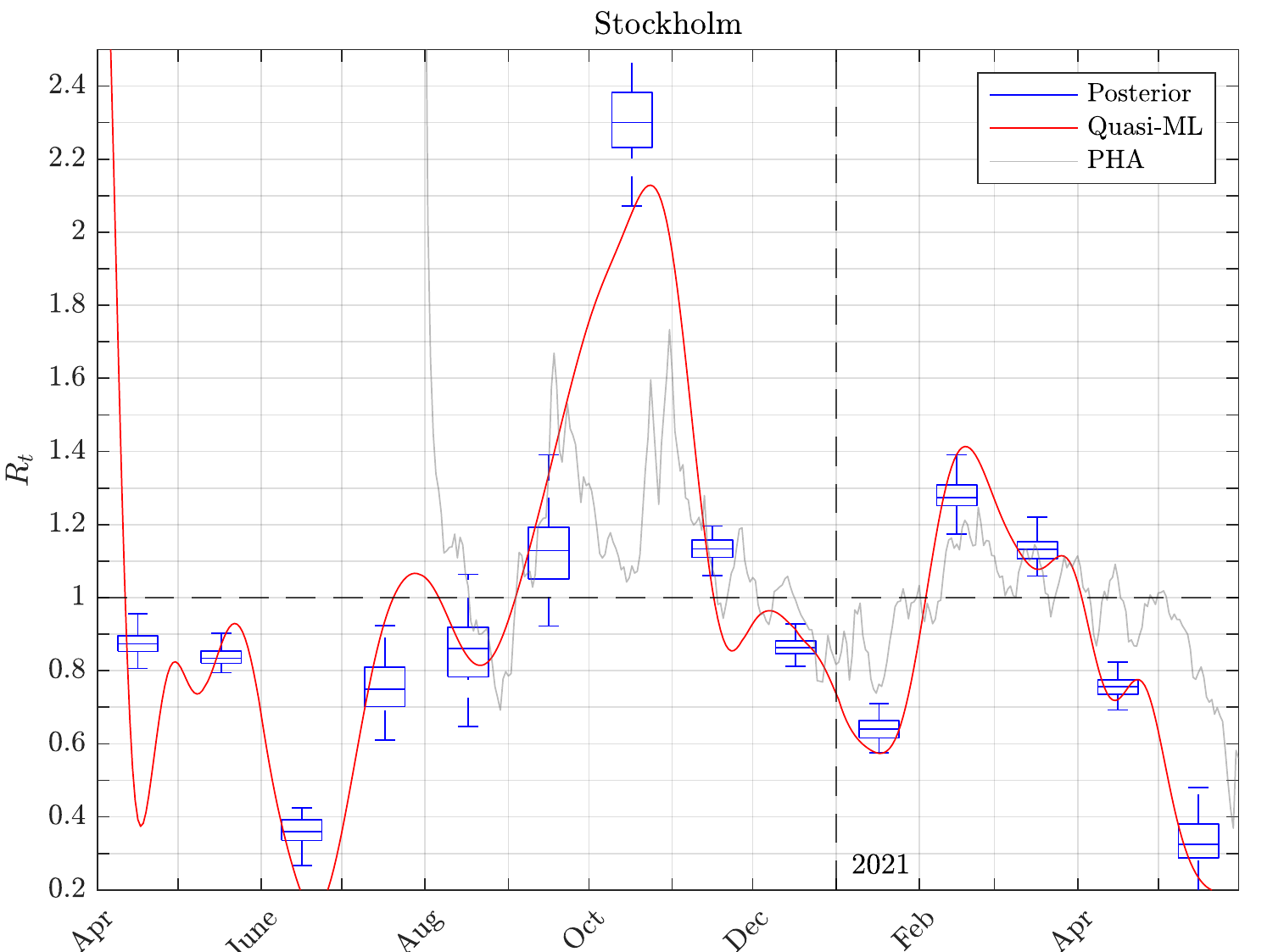}
  \end{subfigure}
  \hfill
  \begin{subfigure}{0.3\textwidth}
    \includegraphics[clip=true, trim = 2.8cm 9.3cm 2.8cm 9.0cm,
    width=1\linewidth]{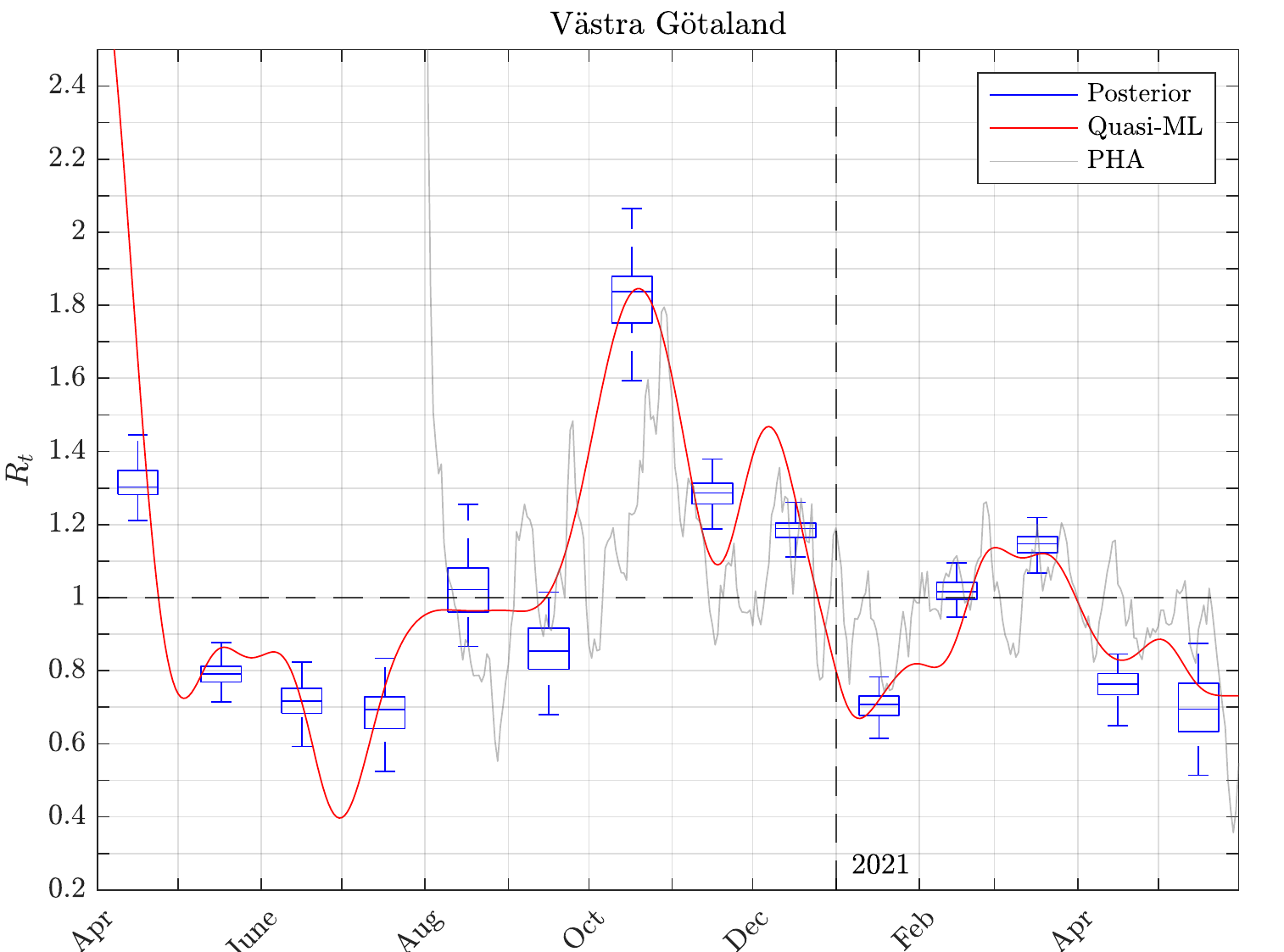}
  \end{subfigure}
  \hfill
  \begin{subfigure}{0.3\textwidth}
    \includegraphics[clip=true, trim = 2.8cm 9.3cm 2.8cm 9.0cm,
    width=1\linewidth]{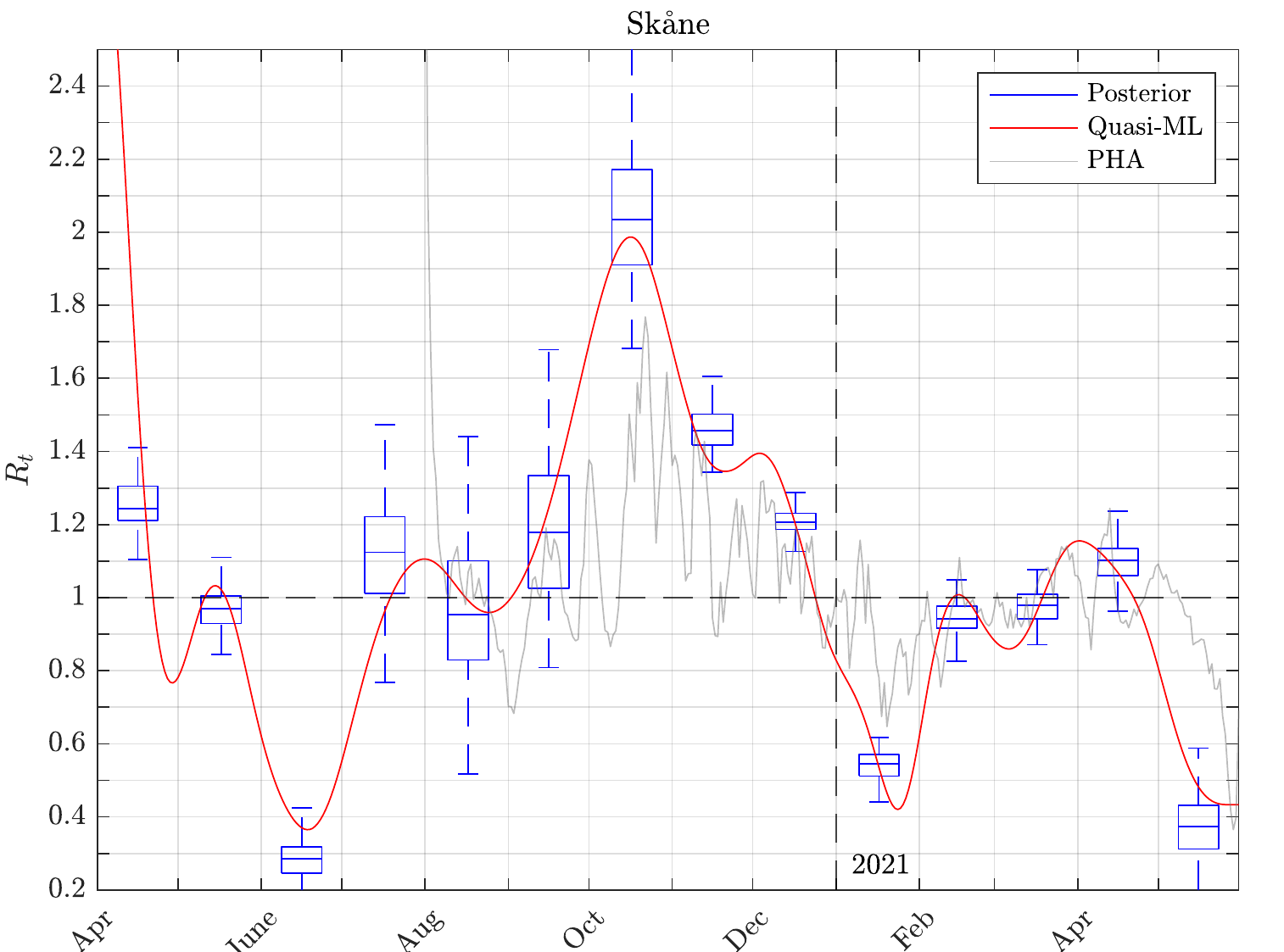}
  \end{subfigure}
  \newline
  \begin{subfigure}{0.3\textwidth}
    \includegraphics[clip=true, trim = 2.8cm 9.3cm 2.8cm 9.0cm,
    width=1\linewidth]{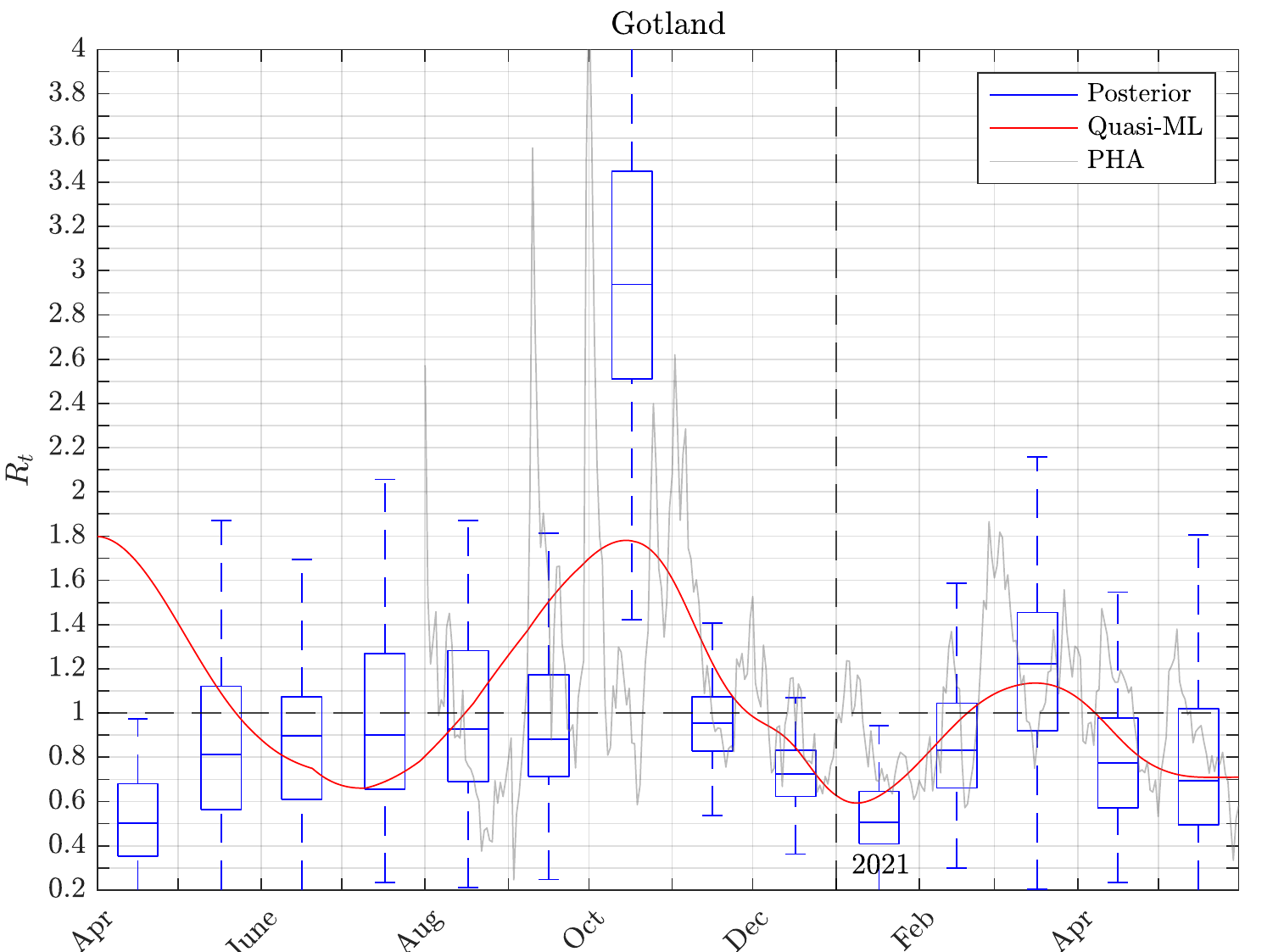}
  \end{subfigure}
  \hfill
  \begin{subfigure}{0.3\textwidth}
    \includegraphics[clip=true, trim = 2.8cm 9.3cm 2.8cm 9.0cm,
    width=1\linewidth]{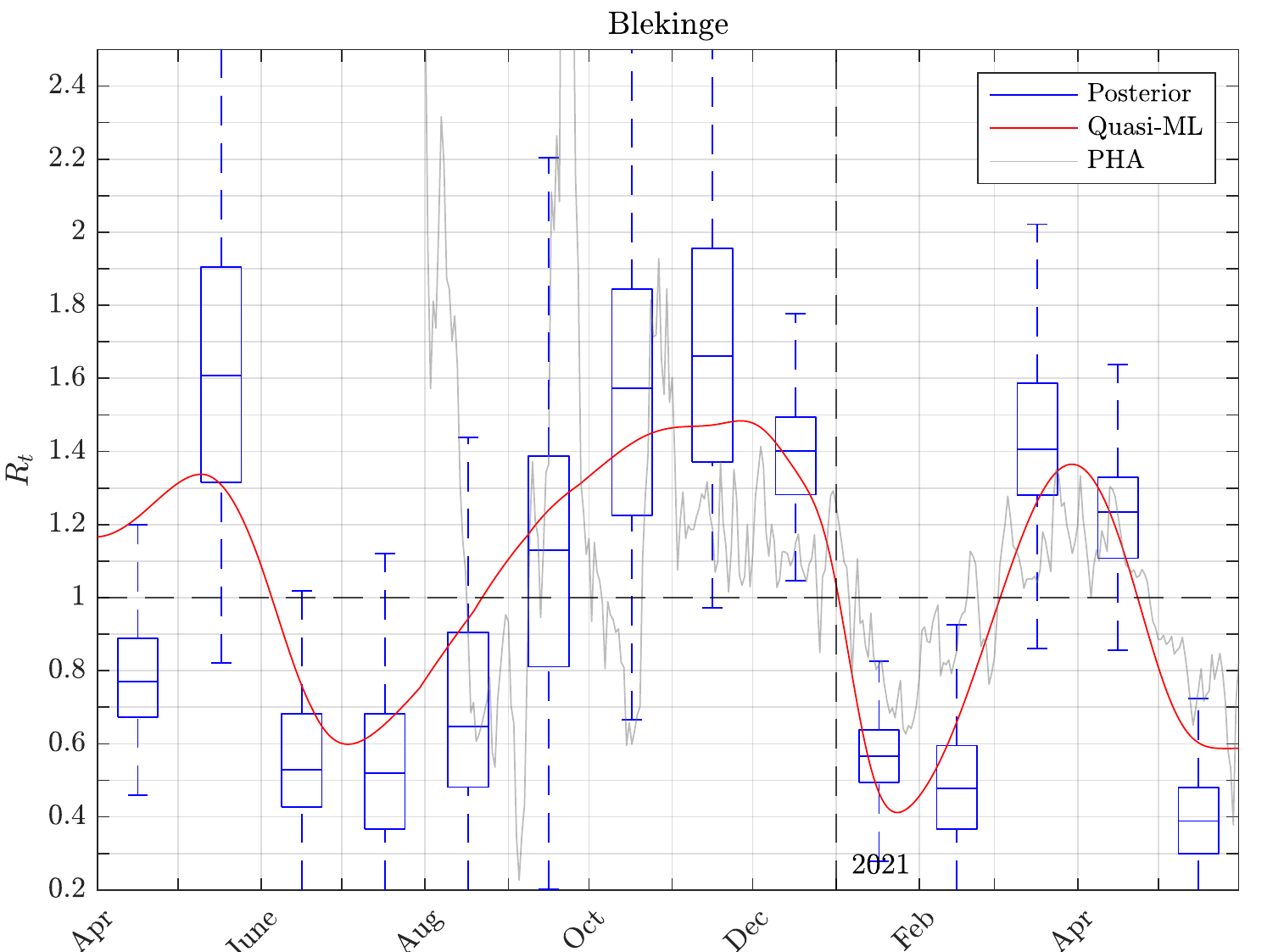}
  \end{subfigure}
  \hfill
  \begin{subfigure}{0.3\textwidth}
    \includegraphics[clip=true, trim = 2.8cm 9.3cm 2.8cm 9.0cm,
    width=1\linewidth]{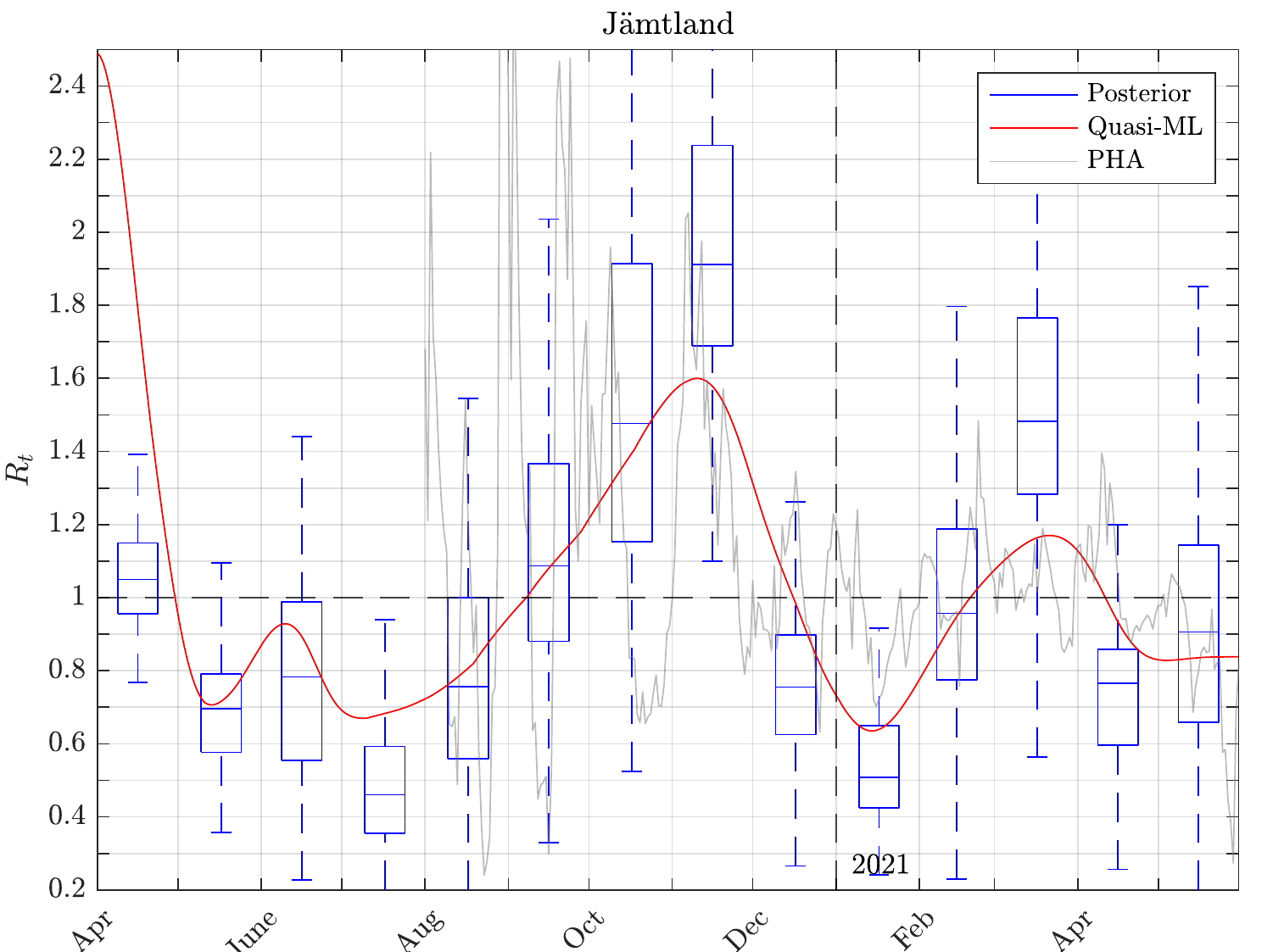}
  \end{subfigure}
  \caption{Our daily reproduction number estimate (in red, with
    box plots for each 4-week constant posterior) and the publicly
    reported PHA estimates from the Swedish screening program (in
    grey). The three largest regions are: Stockholm (top left)
    [\numprint{1864000} total \# of RT-PCR tests administered during
    the 2nd and 3rd waves, August 5, 2020--May 26, 2021], V\"{a}stra
    G\"{o}taland (top center) [\numprint{1570000}], and Sk\aa{}ne (top
    right) [\numprint{1240000}], followed by the estimate for the three
    smallest regions: Gotland (bottom left) [\numprint{48440}],
    Blekinge (bottom center) [\numprint{110300}], and J\"{a}mtland
    (bottom right) [\numprint{184300}]. The standard cost per PCR test
    was \numprint{1400} SEK ($\approx \$150$) \cite{pcr2021skr}.}
  \label{fig:Rposterior_largeNsmall}
\end{figure}

We illustrate the daily $R_t$-estimators for the three largest and the
three smallest (by population) regions: Stockholm, Västra Götaland,
Skåne (largest), and, respectively, Gotland, Jämtland, and Blekinge
(smallest) in Fig.~\ref{fig:Rposterior_largeNsmall}. The relative
uncertainty is visibly smaller in the larger regions than in the
smaller ones, a typical small population effect as the data streams
for the smaller regions consist of smaller counts with larger
uncertainties in a relative sense.

\section*{Data}

\begin{figure}[h]\centering
  \includegraphics[clip = true,
  trim = 1cm 2cm 1cm 3cm,
  width=0.75\linewidth]{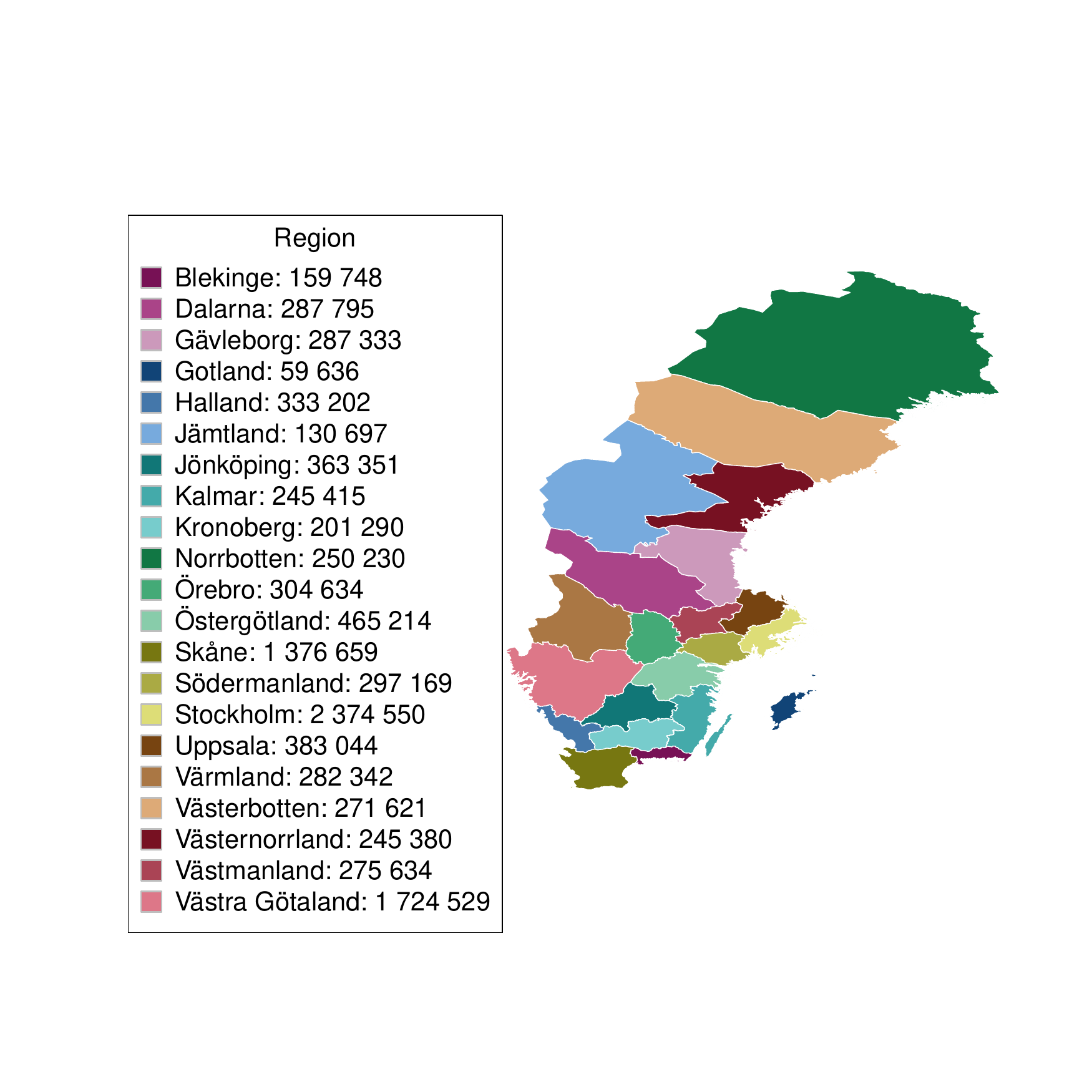}
  \caption{Sweden is divided into 21 regions. We find a posterior
    distribution for each of the individual regions as each region
    publish their own data of observables. In the legend the
    population size of each region is given (as of December 9, 2019
    \cite{data2020scb}), which spans from the smallest: Gotland
    (\numprint{59636}), to the largest: Stockholm
    (\numprint{2374550}).}
  \label{fig:sweden_map}
\end{figure}

The daily observations we use in the inference is retrieved from
\verb;c19.se; using Python scripts. In turn, \verb;c19.se; fetches the
data from the official regional sources on a daily basis. Sweden has
21 regions (Fig.~\ref{fig:sweden_map}) that are in charge of providing
for the healthcare of its population. They also present data on the
number of hospitalized patients, the number of patients in intensive
care, and the daily number of diseased by COVID-19. We use a couple of
other datasets for prior generation and for posterior validation, see
Tab.~\ref{tab:datasource}. The sets \verb;C19; and NBHW$_D$ had double
use, however, the prior information obtained from these were on an
aggregated level.

\begin{table}[h]
  \centering
  {\small
    \begin{tabular}{lrrrrr}
      \hline
      Name       & Description & Time-series & Regional & Use & Source \\
      \hline
      \verb;C19; & Collection of regional reports $(H,W,D)$ & Yes & Yes & Prior, Inference & \cite{data2020c19} \\
      PHA     & Incidence of confirmed cases $(I_{\text{inc}})$ & Yes & Yes & Validation  & \cite{data2020fhm} \\
      NBHW$_D$    & Total \# of deaths  $(D_I,D_H,D_W)$  & No & No  & Prior, Validation & \cite{ss2020avlid} \\
      NBHW$_\gamma$   & Recovery times at hospital and ICU  & No & No & Prior      & \cite{ss2020vard} \\
      SIR$_D$  & \# of deaths as ICU $(D_W)$          & Yes  & Yes &  Prior            & \cite{sir2020mort} \\
      FHM$_{R_t}$ & Estimated $R_t$ per region from $I_{\text{inc}}$ & Yes & Yes & Validation & \cite{rtal2021fhm} \\
      FHM$_{Rec}^1$ & PHA seroprevalence study, outpatient care & No &
                                                                       No & Validation & \cite{fhm2021pavisning} \\
      FHM$_{Rec}^2$ & PHA seroprevalence study, blood donors & No & No & Validation & \cite{fhm2021pavisning}\\
      Castro$_R$ & Estimated Recovered in Stockholm & Yes & No & Validation & \cite{castro2021seropositivity} \\
      SCB$_{\text{trans}}$ & Regional in-/out commuting & No & No & Prior & \cite{pend2019scb} \\
      \hline
    \end{tabular}
    \caption{Summary of data sources used for prior, for inference,
      and for validation.}
    \label{tab:datasource}
  }
\end{table}

\subsection*{Data pre-processing}

Our main data sources were updated each day and incorrect or
questionable data points were common. Two errors that need to be
addressed are next described.
First, there are impossible or very extreme updates, e.g., a
negative incidence or very large jumps, most likely due to an
accumulation of delayed reports, or in some cases possibly in an
attempt to correct earlier reports. For example, a delay in the
reporting of deceased has been described in the Swedish data set
\cite{altmejd2020nowcasting}, and was there explained as ``batch''
reporting. In similar spirit, there are also a few missing data
points. However, these are all found very early on in the data history
and could safely be replaced with \verb;NaN;-values, thus simply
ignored.
Second, and more problematic from a model-based perspective, is the
fact that parts of the reporting display a strong periodic component,
which is not supported by the model and rather most likely an
effect of weekday periodicity.

\begin{SCfigure}
  \includegraphics[clip = true, trim =
  4.5cm 9.8cm 5.0cm 9.5cm,width=0.5\textwidth] {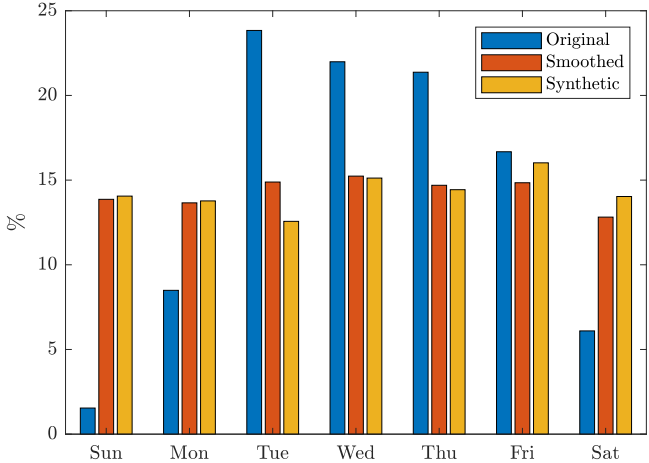}
  \caption{Reports of deceased per weekday over the period April 1,
    2020--May 31, 2021 for the Stockholm region. The smoothing
    procedure attempts to correct for weekly reporting patterns in the
    original data and behaves similarly to a bootstrap replicate for
    the same time period and region.}
  \label{fig:smoothing}
\end{SCfigure}

The pre-processing of data is done in two steps. First, we ensure that
there is no negative incidence data. Negative values are detected and
set to 0, with values backwards in time corrected in a `reasonable'
way using a simple linear decaying memory kernel, and with the
corresponding cumulative compartment adjusted accordingly. In this
first step we also replace missing or manually found early outlier
measurements with either linearly interpolated data from adjacent
days, or using \verb;NaN;-values.
The second step involves smoothing the cumulative fields and the
incidence fields to mitigate the effects of weekly periodic lags or
batch reporting. The smoothing algorithm removes unlikely incidences
by starting from the most extreme outliers, e.g.,
$\ge [10,9,\ldots,2]$ standard deviations under a Poissonian
approximation, and spreads them out during the period before in such a
way as to dampen weekday dependencies in reporting. Most weekly
periodicity is removed by this procedure, see Fig.~\ref{fig:smoothing}
for an example, and whatever remains does not seem to interfere with
the Bayesian inference. The total effect the pre-processing has on the
data can be measured in the sense of the mean maximum relative
difference,
\begin{align}
  \label{eq:minmaxdiff}
  d_{\text{smooth}} &= N_t^{-1} \sum_t \max_i
                       \frac{|X^{(i)}_{\text{smooth}}-X^{(i)}|(t)}{\max(1,|X^{(i)}(t)|)},
\end{align}
where $i \in \{1,2,3\}$ for compartment $X \in [H,W,D]$, i.e., the
three data compartments and over some period of time of length
$N_t$. We report this measure for the period April 1, 2000--May 31,
2021 in Tab.~\ref{tab:bootstrap}. The population weighted national
average difference is $6.6\%$.

\section*{Compartment model}

\begin{figure}[tbhp] \centering
  \includegraphics[width=0.7\textwidth]{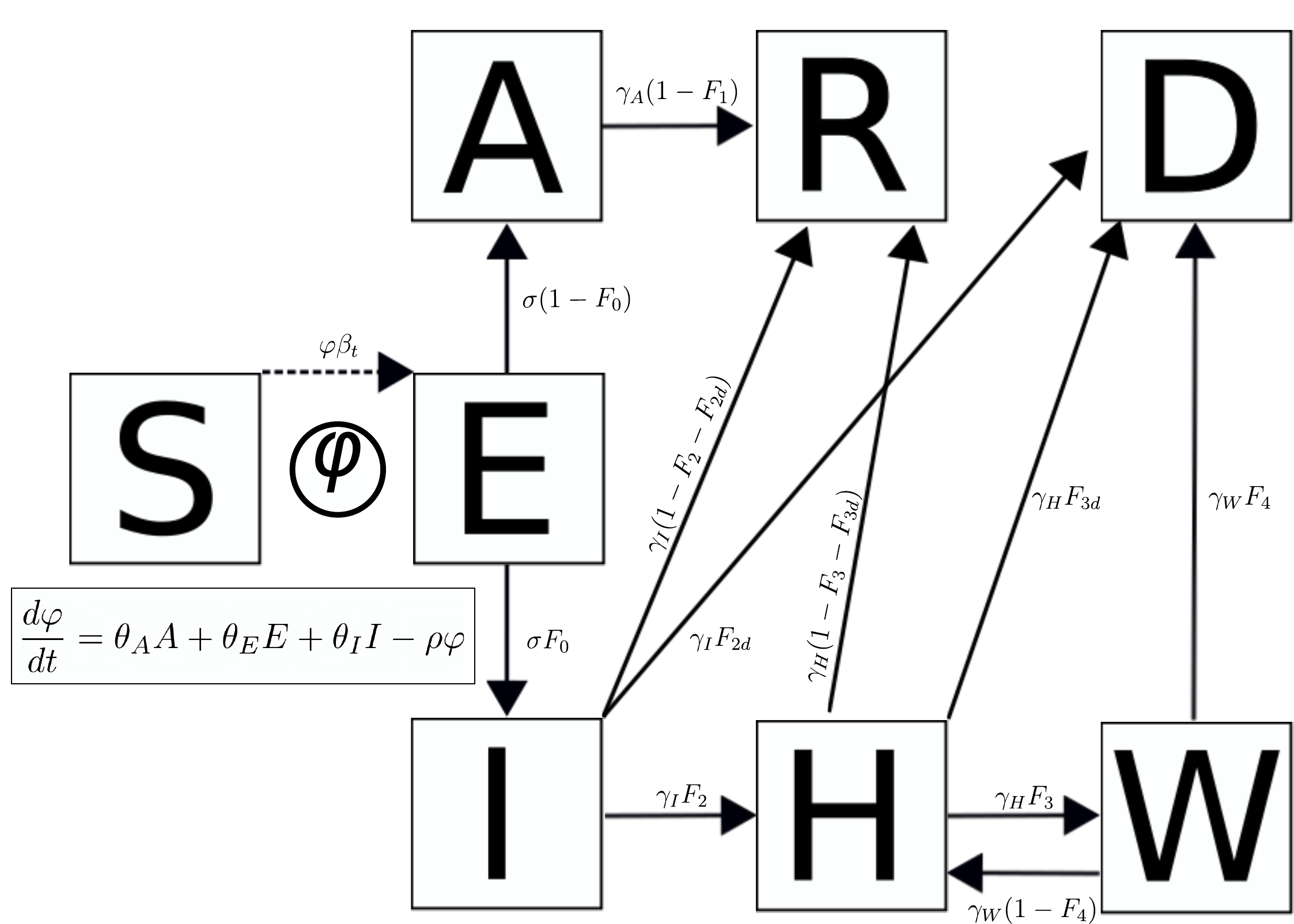}
  \caption{Detailed compartment model of Fig.~\ref{fig:compartments}
    with related transition rates. The dashed arrow indicates the
    interaction with the environmental compartment $\varphi$.}
  \label{fig:compartments_detailed}
\end{figure}

The extended SEIR model is depicted in
Fig.~\ref{fig:compartments_detailed}. The transitions in the figure
are modeled via \emph{transition rates} corresponding to exponentially
distributed waiting times in a continuous-time Markov chain
interpretation.

The compartment model we consider stems from the
SIS$_{\text{E}}$-model
\cite{FreeLivingInfectiveStages}(Chap.~11). This model contains three
state variables, $[S,I,\varphi]$. The first two are integer
compartments and are defined by the Markovian transitions
\begin{align}
  \label{eq:SIS}
  S &\xrightarrow{\beta \varphi \Sigma^{-1}} I, \qquad
      I \xrightarrow{\gamma} S,
\end{align}
with $\Sigma = S+I$, the total population size. Susceptible
individuals turn infected at a rate proportional to the local
infectious pressure $\varphi$, and infected individuals recover at a
rate $\gamma$. The last compartment $\varphi$ represents the
environmental pathogen load and follows the dynamics \cite{siminf_Ch}
\begin{align}
  \label{eq:_E}
  \varphi'(t) &= \theta_I I(t) -\rho \varphi(t).
\end{align}
Infected individuals shed the pathogen into the environment, thus
sourcing the infectious pressure, which also decays at a fixed rate
$\rho$.

\review{The $R_0$-number for the SIS$_{\text{E}}$-model is given by
  $R_0 = \theta_I\beta/(\gamma\rho)$ provided that $\varphi$ is not
  considered a \emph{state-at-infection} \cite{Diekmann2010}, or
  alternatively, $R_0^{(\varphi)} = \sqrt{R_0}$ if this interpretation
  is more natural, e.g., for vector-borne diseases
  \cite{VandenDriessche2017}. In either case $\theta_I = \rho$ is a
  convenient non-dimensionalization.}

Our extended SEIR-model involves two types of parameters: \emph{rates}
and \emph{fractions}. The rates are $[\beta_t$, $\sigma$, $\gamma_I$,
$\gamma_A$, $\gamma_H$, $\gamma_W$, $\theta_A$, $\theta_E$,
$\theta_I$,$\rho]$, and the fractions $[F_0$, $F_1$, $F_2$, $F_{2d}$,
$F_3$, $F_{3d}$, $F_4]$.  We may further divide the rates into waiting
times and those governing the infectious pressure. The waiting times
are $\sigma$, $\gamma_I$, $\gamma_A$, $\gamma_H$, and $\gamma_W$, and
are understood as the inverse of the mean time an individual stays in
a certain compartment, e.g., the mean recovery time for a symptomatic
infected is $\gamma_I^{-1}$.
The transition from $S$ to $E$ depends on $\varphi$ and $\beta_t$.
The infectious pressure $\varphi$ is sourced by the viral shedding
from asymptomatic $\theta_A$, from exposed $\theta_E$, and from
symptomatic individuals $\theta_I$, and it decreases by the viral
decay rate $\rho$. We use the non-dimensionalization $\theta_I = \rho$
and the scaled variables $\theta_{E} = \theta_{E^*} \times \rho$ and
$\theta_{A} = \theta_{A^*} \times \rho$.

For this model the reproduction number can be determined \review{using
  the next generation method, and under the interpretation that
  $\varphi$ is not a state-at-infection \cite{Diekmann2010},}
\begin{align}
  \label{eq:R_0}
  R_0 = \beta_0  \bigg(
  \frac{\theta_{E^*}}{\sigma} +
  \frac{(1-F_0)\theta_{A^*}}{\gamma_A} +
  \frac{F_0 + (1-F_0)F_1}{\gamma_I}\bigg),
\end{align}
and an identical relation holds for $(R_t,\beta_t)$. \review{As with
  the SIS$_{\text{E}}$-model, we have that
\begin{align}
  \label{eq:R_0phi}
  R_0^{(\varphi)} = \sqrt{R_0}
\end{align}
if instead, $\varphi$ \emph{is} understood as a state-at-infection.}

The fraction parameters determine the probabilistic fates of
individuals in a compartment with more than one exit. They were
determined from demographic averages since the daily data did not
contain the level of detail to support age-dependency (cf.~the
discussion in \matmet).

\begin{align}
  E \rightarrow I: &\quad F_0 = E_2I,\\
  A \rightarrow I: &\quad F_1 = A_2I \quad(=0\, \text{ in our simulations}),\\
  I \rightarrow H: &\quad F_2 = \text{HOSP},\\
  H \rightarrow W: &\quad F_3 = \text{IC~HOSP}, \\
  H \rightarrow D: &\quad F_{3d} = \text{SIR MORT } \times \text{ HOSP
                     MORT}, \label{eq:HOSPMORT} \\
  W \rightarrow D: &\quad F_4 = \text{SIR MORT}, \quad \mbox{ and,} \label{eq:SIRMORT} \\
  I \rightarrow D: &\quad F_{2d} = \max\left(0, \frac{\text{IFR}}{E_2I} - \frac{(\text{HOSP MORT} +
                     \text{IC~HOSP}) \times \text{SIR MORT} \times \text{HOSP}}{1 -
                     \text{IC~HOSP} \times (1 - \text{SIR MORT})}\right). \label{eq:IMORT}
\end{align}
The final relation is best explained from the discussion leading to
\eqref{eq:IFRrelIMORT} below. The remaining fractions of the model can
generally be obtained by requiring that the total sum of all outgoing
fractions is one. That is, the fraction which recovers from
compartment $X$ is given by $1 - F_X - F_{Xd}$, where $F_X$ is the
fraction that enters the next state in the chain and where $F_{Xd}$ is
the fraction that dies.

Note that the model parameters covering fatalities,
$(F_{2d},F_{3d},F_4)$, are obtained from the more natural parameters
$(\text{SIR MORT}, \text{HOSP MORT},\text{IFR})$ via
Eqs.~(\ref{eq:HOSPMORT})--(\ref{eq:IMORT}).

\subsection*{Network effect for national scale simulations}

We did not find that connecting the regions improved the fit to data
and, in the end, we therefore decided not to use this technique on a
regular basis. However, since we did use it in
Tab.~\ref{tab:PastWeekly} and for completeness, we describe below how
it was implemented.

The network connection is obtained by introducing a prior commuting
intensity factor $\lambda$ that allows for connections between the
regions. This affects only the calculations on a national level, and
it does so mainly by adding some correlation between connected
regions.

The network is defined by a connection matrix $D$ which is a 21-by-21
square matrix with a zero diagonal. For the $i$th row, each column $j$
contains the proportion of individuals commuting into region $i$ from
region $j$. In turn, the connection matrix itself is found by a linear
programming formulation on the volumes of individuals commuting in and
out per each region \cite{pend2019scb}. At each forward step in the
simulation, the infectious pressure $\varphi_k = \varphi(t_k)$ is then
updated as
\begin{align}
  \varphi_{k+1} &= \bar{\varphi}_{k+1}+\lambda (D\varphi_k
                  - d \odot \varphi_k),
\end{align}
where $\bar{\varphi}_{k+1}$ is the update according to
Eqs.~(\ref{eq:kalman1})--(\ref{eq:kalman3}), i.e., without any network
effects, and where $d$ is the column sum of $D$ and $\odot$ is
element-wise multiplication.


\section*{Priors}

The prior knowledge relied upon to construct priors stems from several
sources. Below we briefly comment on the techniques and assumptions
used to derive our priors; effective summaries are found in
Tabs.~\ref{tab:prior_stat} and \ref{tab:prior_dynamic}. The process of
constructing priors was ongoing during the whole period fall 2020 to
late spring 2021. We revised the priors whenever we became aware of
new research, when more statistics became available, or we deemed
model tweaks necessary. The final priors (arrived at towards the end
of May 2021) are what we discuss below.

\begin{description}
\item[$\boldsymbol{\sigma}$ and
  $\boldsymbol{\gamma_I}/\boldsymbol{\gamma_A}$] The mean incubation
  time $\sigma^{-1}$ is set to $6.2$ days with support in $[5.4,7]$,
  derived from the mean and 95\% CI given in
  \cite{dhouib2021incubation}. The recovery time for the symptomatic
  infectious $\gamma_I^{-1}$ is less known and we therefore set the
  mean to $7$ days \cite{byrne2020inferred}\cite{ferretti2020quantifying}
  with a wider support of $[4,10]$ days. We assume similarity between
  asymptomatic and symptomatic cases by setting $\gamma_A = \gamma_I$,
  which covers the priors suggested in \cite{byrne2020inferred}.

\item[$\boldsymbol{\gamma_H}$ and $\boldsymbol{\gamma_W}$] We
  determine the priors for hospital recovery times, $\gamma_H^{-1}$
  and $\gamma_W^{-1}$ from a dataset published by the NBHW
  \cite{ss2020vard}, including the distribution of exit times for
  hospital patients with and without ICU care. The dataset is
  right-tailed censored for waiting times over 30 days, which affects
  the ICU patient records. For $\gamma_H^{-1}$, we first make a
  Bayesian fit of an exponential distribution to hospital caring time
  data ($N = \numprint{40507}$ patients) and place a somewhat broader
  beta distribution across the credible interval for the hyper
  parameter thus found, this results in a beta distribution of mean
  $8.9$ and support on $[8.7,9.1]$ [days].  The recovery time under
  intensive care, $\gamma_W^{-1}$, is more complicated since the
  available data involves also non-ICU caring times. Assuming the
  previously determined exponentially distributed waiting times under
  hospital care, we first subtracted two such waiting times (to model
  pre- and post-ICU care, respectively), next made a Bayesian
  exponential fit to the remaining time ($N = \numprint{4039}$
  patients). This fit was judged a bit worse than for non-ICU care
  time and so we used a larger enclosing support of $[10.8,13.9]$ with
  mean $12.2$ [days].

\item[$\boldsymbol{E_2I}$] The prior for the fraction
  $E_2I \, (E \rightarrow I)$ was taken from
  \cite{alene2021magnitude}: $0.75\, (0.62, 0.84)$ (95\% CI), for
  which we determine a scaled beta distribution that fits the mean and
  the given CI interval. The resulting prior distribution has mean
  $0.75$ and support on $[0.014, 1]$.

\item[IFR] The IFR, i.e., the eventual fraction of infected
  individuals that dies ($E \rightarrow D$ in our model), is volatile
  in that the IFR depends strongly on the age distribution of the
  region as well as on the quality of the health care and the
  currently dominating virus variant \cite{brazeau2020report,
    fhm2020IFR}. We therefore settled on a scaled beta distribution
  with positive skewed and rather wide support: $0.67\%, [0, 2]\%$.

\item[HOSP~MORT and SIR~MORT] The priors for the mortality at hospital
  and intensive care (HOSP~MORT, SIR~MORT) are recovered directly from
  mortality datasets linked to hospitalization deaths
  \cite{ss2020avlid} ($N = \numprint{46236}$ patients and
  \numprint{5729} diseased) and ICU deaths \cite{sir2020mort}
  ($N = \numprint{5744}$ patients and \numprint{1357} diseased), and
  are kept as fixed constants.

\item[IC~HOSP] To find our prior for the proportion of hospitalized
  patients needing intensive care, we use the \verb;c19.se; data in
  aggregate form and extract the percentiles ($[0.5, 50, 99.5]\%$) of
  the quotient $[H:W]$ between the number of hospitalized and ICU
  patients for all data available. We next assume a relation of the
  form
  \begin{align}
    H &\sim [H:W] \times W, \\
    \intertext{for $W$ approximately stationary and thus satisfying a
    balance condition. This means that}
    W_{\text{in}} &= \gamma_H \times H \times \text{IC~HOSP} \approx
                    \gamma_W \times W =
                    W_{\text{out}}. \\
    \intertext{It then follows that}
    \text{IC~HOSP} &\approx \gamma_W W/(\gamma_H H) \sim
                     \gamma_W/(\gamma_H [H:W]).
  \end{align}
  Transforming the percentiles of $[H:W]$ accordingly, we then find a
  beta distribution with support on these ($[0.065, 0.95]$) and a mean
  at the median ($0.18$).

\item[HOSP] At this point we have the relation
  \begin{align}
    \text{IFR} &= [E_2I] \times
    \text{I~MORT}+[H] \times
    \text{HOSP~MORT}  \times
    \text{SIR~MORT}+[W]\times \text{SIR~MORT},
  \end{align}
  in terms of the asymptotic fractions infected with symptoms
  $[E_2I]$, under hospital care $[H]$, and under intensive care $[W]$,
  and all in relation to the exposed population $E$. We find the
  asymptotic fractions $[H]$ and $[W]$ by considering the dynamics
  expressed in Fig.~\ref{fig:compartments_detailed} and recognizing a
  geometric series,
  \begin{align}
    [H] &= E_2I \times \text{HOSP}\times(1+x+x^2+\dots)
          =  E_2I\times \text{HOSP}/(1-x),
  \end{align}
  with $x = \text{IC~HOSP}\times(1-\text{SIR~MORT})$. Similarly,
  \begin{align}
    [W] &=
          E_2I\times\text{HOSP}\times\text{IC~HOSP}\times(1+x+x^2+\dots)
          = E_2I\times\text{HOSP}\times\text{IC~HOSP}/(1-x).
  \end{align}
  We thus arrive at the relation
  \begin{align}
    \label{eq:IFRrelIMORT}
    \text{IFR} &=  E_2I \times [\text{I~MORT}+(\text{HOSP~MORT}+
                 \text{IC~HOSP})\times\text{SIR~MORT}
                 \times\text{HOSP}/(1-x)].
  \end{align}
  We next find a prior for the fraction of symptomatic individuals
  that enters the hospital (HOSP) as follows. The model in itself
  restricts the dead source compartments to $I$, $H$, and $W$. By the
  previous calculations the total risk of death from compartment $I$
  can be decomposed into
  \begin{align}
    \nonumber
    \text{CFR} &= \text{I~MORT}+\underbrace{\left(
                 \text{HOSP~MORT}  \times
                 \text{SIR~MORT} \times \text{HOSP}/(1-x) \right)}_{=: \text{H~MORT}}
                 +\\
    &+\underbrace{\left( \text{SIR~MORT} \times
                 \text{HOSP}\times\text{IC~HOSP}/(1-x)\right)}_{=: \text{W~MORT}}.
  \end{align}
  To close the system of equations we assume the relation
  \begin{align}
    \text{I MORT} &\sim [I:HW] \times (\text{H~MORT}+\text{W~MORT}),
  \end{align}
  for some unknown scaling $[I:HW]$. Connecting with aggregated
  mortality data for total deaths from \verb;c19; and hospital deaths
  from \cite{ss2020avlid}, we find $[I:HW] \approx 0.98$ which we
  simply take to be a beta distribution with support $[0,2]$ and mean
  $1$. At this point we resort to a direct Monte Carlo simulation of
  \eqref{eq:IFRrelIMORT} and find samples from $\text{HOSP}$ using
  \begin{align}
    \text{HOSP} &= \frac{\text{IFR}/E_2I}{(1+[I:HW]) \times
                  \left( \text{HOSP~MORT}+\text{IC~HOSP} \right)
                  \times \text{HOSP}/(1-x)}
  \end{align}
  We fit a scaled beta distribution to \numprint{100000} samples from
  this distribution and finally obtain a beta distribution with mean
  $0.033$ and support $[0, 0.17]$.

\item[$\boldsymbol{\theta_E}$ and $\boldsymbol{\theta_A}$] The viral
  shedding from compartments $E$ and $A$, $\theta_{E^*}$ and
  $\theta_{A^*}$, respectively, is assumed to be uncertain but
  bounded. We assign the same prior to both shedding rates, a scaled
  beta distribution with the mode at 1 and support in $[0, 2]$.

\item[Infectious half-life $\boldsymbol{\tau_{1/2}}$ and
  $\boldsymbol{\rho}$] The decay rate $\rho$ in $\varphi$ is defined
  as $\rho= \log(2)/\tau_{1/2}$ for $\tau_{1/2}$ the infectious
  half-life. The prior for the latter is taken to be uniform between 1
  and 12 hours, realized as a uniform distribution between $1/24$ and
  $12/24$. This encloses the estimate from \cite{van2020aerosol} which
  suggests 3 hours.

\item[$\boldsymbol{\lambda}$] As mentioned, we do not find the
  posterior for the network coupling $\lambda$. But we do define a
  distribution which we keep fixed and sample from when performing
  some of our Sweden-level simulations. We assume a scaled beta
  distribution with mean $\lambda_0$ and support
  $\lambda_0 \times [0.5, 1.5]$, where $\lambda_0 = 8/24 \times
  5/7$. This scaling is meant to achieve the proportion of time at
  work under normal (non-pandemic) circumstances.

\item[$\boldsymbol{R_t}$] Lastly we have $\beta_t$ which we find from
  sampling $R_t$ and using the map \eqref{eq:R_0}. \review{We have
    already noted that the reproduction number depends on the
    interpretation of the state $\varphi$. We obtain a generous prior
    by placing the prior on $R_0$ found in
    \cite{alimohamadi2020estimate} at $R_t^{(\varphi)}$; a truncated
    lognormal distribution with log mean $\log(1.3)$ and log standard
    deviation $0.4$, and with support $[0, 4]$. The prior for $R_t$ is
    then simply the square of this, see Tab.~\ref{tab:prior_dynamic}.}
\end{description}

\begin{table}
  \centering
  {\small
    \begin{tabular}{lrrrrr}
      \hline Parameter & Prior & Mean & Description & Empirical & Source\\
      \hline
      \textbf{Inferred} & & & & & \\
      $\sigma$ & $\mathcal{B}_{[0.14, 0.19]}(2, 2.6)$ &$6.2^{-1}$ & Incubation period &--  & \cite{dhouib2021incubation}  \\
      $\gamma_I$ & $\mathcal{B}_{[0.1,0.25]}(2,5)$         & $7.0^{-1}$ & Infectious period & --  & \cite{byrne2020inferred}\cite{ferretti2020quantifying}\\
      $\gamma_H$ & $\mathcal{B}_{[0.110,0.114]}(3, 3)$    &   $8.9^{-1}$ & Hospital period & Yes &\cite{ss2020vard}\\
      $\gamma_W$ & $\mathcal{B}_{[0.072,0.092]}(2, 2)$    & $12^{-1}$& Intensive care period & Yes & \cite{ss2020vard}\\
      $E_2I (E \rightarrow I)$  &$\mathcal{B}_{[0.014,1]}(52.56,17.85)$&   $75\%$      & Fraction $E \to I$ & --   &   \cite{alene2021magnitude}\\
      $\text{IC~HOSP} (H \rightarrow W)$&$\mathcal{B}_{[0.065,0.94]}(2,13.21)$  & $18\%$ &Fraction $H \to W$ & Yes  & \cite{data2020c19}\\
      $\text{HOSP} (I \rightarrow H)$&$\mathcal{B}_{[0,0.17]}(2.03,8.28)$ &$3.3\%$ &Fraction $I \to H$ & Yes & \cite{ss2020vard}\cite{data2020c19}\\
      $\theta_{E^*}$ & $\mathcal{B}_{[0, 2]}(2,2)$  & 1.0 & Source  $E\to\varphi$  & -- & This paper\\
      $\theta_{A^*}$ & $\mathcal{B}_{[0, 2]}(2,2)$ & 1.0 &  Source $A\to\varphi$ & -- & This paper\\
      $\tau_{1/2}$ & $\mathcal{U}(1/24,12/24)$          & $6.5/24$ & Infectious pressure half-life
                                                    &--  & \cite{van2020aerosol} \\ \hline
      \textbf{Not inferred} & & & & & \\
      $\gamma_A$ & $\gamma_I$  & $7.0^{-1}$ & Infectious period ($A \to$)   &--  & \cite{byrne2020inferred}\\
      $(\text{HOSP MORT}) H \rightarrow D$ & $0.1322$      & $\leftarrow$ & Mortality risk from $H$ & Yes & \cite{ss2020avlid}\cite{sir2020mort}\\
      $(\text{SIR MORT})W \rightarrow D$ & $0.2129$        & $\leftarrow$ & Mortality risk from $W$ & Yes & \cite{sir2020mort}\cite{straalin2021mortality}\\
      $\rho$ & $\log(2)/\tau_{1/2}$ & $\leftarrow$ &
                                                               Infectious pressure decay &  -- & This paper\\
      $\lambda_0$ & $8/24 \times 5/7$& $\leftarrow$ & Commuting intensity & -- & This paper\\
      $\lambda$ & $\lambda_0 \times \mathcal{B}_{[0.5,1.5]}(2, 2)$& $\lambda_0$ & Commuting intensity & -- & This paper\\
      $I \rightarrow D$ & Eq.~\eqref{eq:IMORT} & $0.24\%$ &    Implicitly defined & -- & This paper\\
      \hline
    \end{tabular}
    \caption{Prior distributions for the \textit{static} parameters. \review{The
        notation $\mathcal{B}_{[L,U]}(a,b)$ denotes a beta distribution of
        parameters $(a,b)$, scaled and shifted into the interval
        $[L,U]$.}}
    \label{tab:prior_stat}
  }
\end{table}

\begin{table}[h]
  \centering
  \begin{tabular}{lrrrrr} \hline Parameter & Prior & Mean & Std & Description & Source\\
    \hline
    IFR   & $\mathcal{B}_{[0, 0.02]}(2, 4)$ & $0.67\%$ & $0.36\%$ & Infection fatality rate  & \cite{brazeau2020report} \\
    \review{$R_t$} & $\log\mathcal{N}_{[0, 16]}(\log(1.69),0.8)$ &
                                                                   $2.3$ & $2.0$ &
                                                                   Reproduction number & $\equiv (R_t^{(\varphi)})^2$\\

    \review{$R_t^{(\varphi)}$} & $\log\mathcal{N}_{[0,
                                 4]}(\log(1.3),0.4)$ & $1.4$ & $0.57$ & Eqs.~\eqref{eq:R_0}--\eqref{eq:R_0phi} & \cite{alimohamadi2020estimate} \\

    \hline
  \end{tabular}
  \caption{Prior distributions for the \textit{dynamic}
    parameters. \review{The notation
      $\log\mathcal{N}_{[0,U]}(\mu,\sigma)$ denotes a truncated
      lognormal distribution with associated normal distribution
      parameters $(\mu,\sigma)$, truncated to the interval $[0,U]$.}}
  \label{tab:prior_dynamic}
\end{table}

\subsection*{Prior predictive}

\begin{figure}[tbhp] \centering \includegraphics[clip = true,
  trim = 4.5cm 9.3cm 4.5cm 9.5cm,
  width=0.7\linewidth]{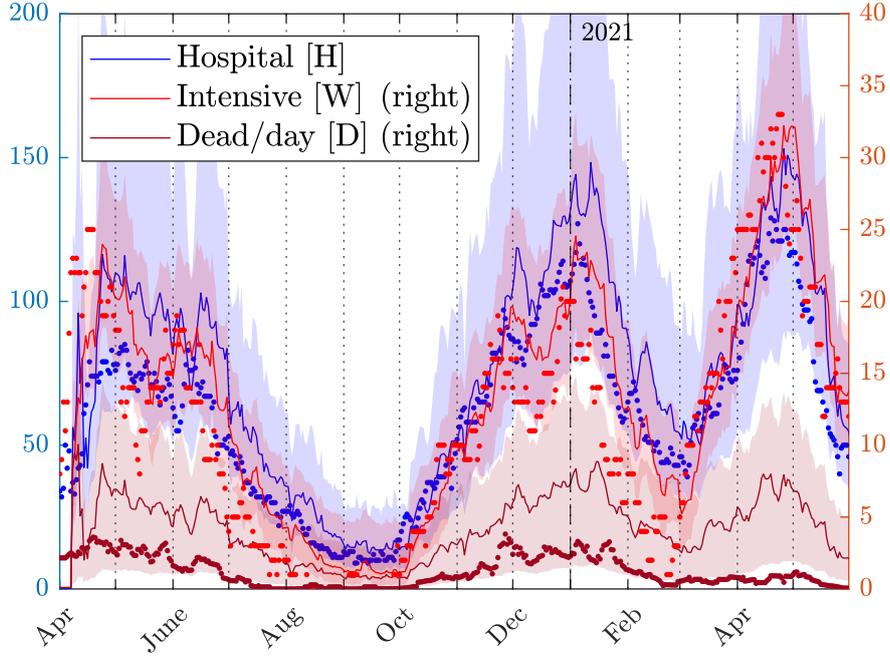}
  \caption{7-day ahead prediction with 68\% CrI (shaded) using
    \numprint{1000} samples from the prior distribution and with data
    for reference (points). See Fig.~\ref{fig:lag_uppsala} for the
    posterior equivalent.}
  \label{fig:prior_pred}
\end{figure}

We assess the quality of the prior distribution through some samples
from the prior predictive distribution. We generate 7-day ahead
predictions using the same set-up as in Fig.~\ref{fig:lag_uppsala} and
illustrate the results in Fig.~\ref{fig:prior_pred}.

\section*{Kalman filter model}

A linear noise approximation of a continuous-time Markov chain (CTMC)
with exponentially distributed waiting times between the compartments
is employed in the Bayesian modeling. Under the assumption that the
susceptible population decreases slowly (in a relative sense) compared
to the other states of the system, i.e., only a small portion of the
total population is becoming infected over short time horizons, the
dependence of the susceptible population on the infection rate is
neglected and instead captured implicitly via the time-dependence of
$(\beta_t,R_t)$. The model is discretized in time, which results in
the linear state-space model
\begin{align}
  \label{eq:kalman1}
  x_{k+1} &=F x_k + w_k,
\end{align}
where $x_k$ is the 8-dimensional state vector consisting of the
compartments
\begin{align}
  \label{eq:kalman2}
  x_k &= \begin{bmatrix}
    I_k & A_k & E_k & \varphi_k & H_k &
    W_k & D_k & R_k
  \end{bmatrix}^\intercal,
\end{align}
$k$ is the time index (daily) and $F$ is given explicitly by
\begin{align}
  \label{eq:kalman3}
  F &= \begin{bmatrix}
    1-\gamma_I & \gamma_A F_1 & \sigma
    F_0 & 0 & 0 & 0 & 0 & 0 \\ 0 & 1-\gamma_A & \sigma (1-F_0) & 0 & 0 & 0
    & 0 & 0 \\ 0 & 0 & 1-\sigma & \beta & 0 & 0 & 0 & 0 \\ 1-e^{-\rho} &
    \theta_A (1-e^{-\rho}) & \theta_E (1-e^{-\rho}) & e^{-\rho} & 0 & 0 &
    0 & 0 \\ \gamma_I F_2 & 0 & 0 & 0 & 1-\gamma_I & \gamma_W (1-F_4) & 0
    & 0 \\ 0 & 0 & 0 & 0 & \gamma_H F_3 & \gamma_W & 0 & 0 \\ \gamma_I
    F_{2d} & 0 & 0 & 0 & \gamma_H F_{3d} & \gamma_W F_4 & 1 & 0 \\
    \gamma_I (1-F_2-F_{2d}) & \gamma_A (1-F_1) & 0 & 0 & \gamma_H
    (1-F_3-F_{3d}) & 0 & 0 & 1
  \end{bmatrix}.
\end{align}
The parameters in this matrix are the ones described in the previous
sections. Note how the equation of state for the infectious pressure
$\varphi$ has been integrated explicitly.

To reduce the transients, we use an initialization of the filter
developed specifically for this purpose. Given the state update matrix
$F_k = F(\beta_k)$, observation matrix $H_k = H$, and time series data
$y = (y_k)$, the first step of the algorithm removes all cumulative
states from $F_0$, $H_0$, and $y_0$ to produce $F_{\text{red}}$,
$H_{\text{red}}$, and $y_{\text{red}}$. The initial non-cumulative
states are then obtained by projecting the dominating eigenvector of
$F_{\text{red}}$ onto the subspace defined by
$H_{\text{red}} x_{\text{red}} = y_{\text{red}}$ under a positivity
constraint. For the \emph{measured} cumulative states, the initial
state itself is taken as the data, while the unmeasured cumulative
states are set to zero. In particular and under reasonable
assumptions, if the system would be initialized from the eigenvector
and simulated without perturbations, the relative magnitude of the
non-cumulative states would remain constant.

An important property of this model is the distribution of the noise
$w_k$. Since a Poisson-distributed transition has a variance
proportional to the number of individuals in the compartment, the
process noise is state-dependent. A term proportional to the squared
compartment population, representing an uncertainty in the transition
``flow'', is also added to the variance, as well as a constant, i.e.,
regularizing, term. Hence for a transition from a state $A$ to state
$B$ with rate $\mu$, the noise covariance matrix is of the form
\begin{equation}
  Q=Q_p + Q_v + Q_0,
\end{equation}
with
\begin{equation}
  Q_p=\begin{bmatrix}
    \mu A & -\mu A \\
    -\mu A & \mu A
  \end{bmatrix},  \quad
  Q_v= \epsilon \begin{bmatrix}
    (\mu A)^2 & 0 \\
    0 & (\mu A)^2
  \end{bmatrix}, \quad
  Q_0=\begin{bmatrix}
    q_A & 0 \\
    0 & q_B
  \end{bmatrix},
\end{equation}
where $\epsilon$, $q_A$ and $q_B$ have positive values. Note that a
negative correlation is induced by the Poisson-noise, while we do not
introduce such terms for $Q_v$ nor $Q_0$. The process noise covariance
of the full model is then calculated by addition of the individual
contributions from all transitions which are encoded in $F$. Note that
the state $\varphi$ is the discretization of an ordinary rather than a
stochastic differential equation, so there are no Poissonian
contributions to the corresponding elements of the covariance
matrix. In our setup, $\epsilon = 0.05^2$ and diagonal elements of
unity for $Q_0$ are chosen, i.e., corresponding to process noise on
the order of single individuals.

The measured signals are given by
\begin{equation} y_k = H x_k + v_k,
\end{equation}
where for measurements of the states $[H,W,D]^\intercal$ we have
\begin{equation}
  H=\begin{bmatrix}
    0 & 0 & 0 & 0 & 1 & 0 & 0 & 0 \\
    0 & 0 & 0 & 0 & 0 & 1 & 0 & 0 \\
    0 & 0 & 0 & 0 & 0 & 0 & 1 & 0
  \end{bmatrix}
\end{equation}
The components of the measurement $v_k$ are assumed to be uncorrelated
and consist of both a constant and a state-dependent term so that the
corresponding covariance matrix becomes
\begin{equation}
  R_k =\begin{bmatrix}
    r_{0,H} + r_{d,H} H_k^2 & 0 & 0 \\
    0 & r_{0,W} + r_{d,W} W_k^2 & 0 \\
    0 & 0 &r_{0,D} + r_{d,D} D_k^2
  \end{bmatrix},
\end{equation}
where the parameter values $r_{0,H} = r_{0,W} = r_{0,D} = 1$ and
$r_{d,H} = r_{d,W} = r_{d,D} = 0.001^2$ are used in our model.

In the Kalman filter, the covariance matrices are calculated based on
state estimates at every iteration. One should, however, note that
optimality results of the Kalman filter only hold when the noise is
Gaussian and additive, i.e., independent of the states, so there is no
theoretical justification for the optimality of the Kalman filter in
the current setup. Calculated results can therefore be understood as
an approximation in density and the Kalman marginal likelihood is an
approximation to the true likelihood. Nonetheless, the filter has
worked rather convincingly in practice.


\section*{Posterior sampling}

When performing Bayesian inference on models with intractable
likelihoods one common approach is Approximate Bayesian Computations
(ABC), also referred to as likelihood-free inference
\cite{marin2012approximate}\cite{sisson2018handbook}. The cornerstone in
ABC is to use a simulator $y \sim F(\theta)$ to generate data. These
generated data are then compared to the observed data and the
``distance'' between them acts as a proxy for the likelihood of the
parameter given the observation.

A flavor of ABC called synthetic likelihoods (SL)
\cite{wood2010statistical}, finds the proxy-likelihood by generating
multiple data samples per parameter proposal, and, assuming asymptotic
normality, computes the then tractable likelihood of the observations.
In our case of using a Kalman filter estimator, the likelihood
estimate is the Kalman marginal likelihood. This likelihood acts in
similar spirit as the SL, since the filter can be viewed as the limit
of multiple simulations under a Gaussian assumption. However, the
analogy is not perfect since the Kalman filter implements correction
steps for each new data point.

We use the Kalman Likelihood in the Adaptive Metropolis algorithm
(``KLAM''). This is similar to the classical Metropolis algorithm but
with an adaptive proposal function \cite{haario2001adaptive}. The
proposal function is a multivariate normal distribution with mean at
the current parameter point and an adjustable covariance matrix
$\mathcal{N}(x_{t-1},C_t)$, where
\begin{align}
  C_t = \begin{cases}
    C_0, &t\leq t_0\\
    s\,\text{cov}(x_0,\ldots,x_{t-1}) + s\varepsilon I_d,  &\text{otherwise}.
  \end{cases}
\end{align}
We assume a diagonal initial covariance $C_0 = 0.001 \times I_d$, for
the $d$-dimensional identity $I_d$, and we start adapting after
$t_0 = 10$ accepted proposals. We also use the step-length tuning
parameter $s = 0.05 \times 2.4^{2/d}$, and we run each region
parameter posterior chain for four parallel $5 \times 10^4$ samples
resulting in a Gelman-Rubin score below $1.1$, e.g., $= 1.01$ for
Uppsala.

By using the Kalman filter likelihood, samples are fast to generate. A
full regional posterior of $2 \times 10^5$ (minus $1 \times 10^4$ as
burn-in) is generated in a little over 30 minutes on an Intel
(4$\times$)Core i7-6820HQ CPU @ 2.70GHz. The regional posteriors are
also solved independently of each other, thus allowing us to sample
them in parallel using Matlab's \verb;parfor; parallel for-loop. A
full national posterior can thus be generated in little over 12
hours. The fast sampling is made possible not only thanks to the
filter set-up, but also thanks to the time-sequential character of the
problem. We repeated the inference each week, and could use the
previous weeks' and regions' posteriors as initial guesses for the new
posteriors.


\section*{Dynamic optimization solution}

With posterior estimates of $\beta_t$ available in four-week
intervals, we now explain further the method to estimate the daily
infection recruitment $\beta_k = \beta(t_k)$ and as a result the daily
reproduction number. The methodology relies on minimizing the negative
logarithmic likelihood. For that purpose, let $B$ denote a vector of
consecutive $\beta_t$'s as
\begin{equation}
  B = \begin{bmatrix}
    \beta_1 &
    \beta_2 &
    \dots &
    \beta_K
  \end{bmatrix}^\intercal,
\end{equation}
and $\tilde B$ denote the corresponding estimate. Writing the
dependence on $B$ explicitly, the optimization problem can then be
formulated as
\begin{equation}\label{eq:fullOpt}
  \tilde B = \argmin_B -\ell^{(K)}(B) + c \Delta B^\intercal \Delta B,
\end{equation}
where $\ell^{(K)}(B)$ denotes the logarithmic marginal likelihood
computed over the time horizon $[1,K]$, i.e.
\begin{equation}
  \ell^{(K)}(B) = -\sum_{k=1}^K
  \frac{1}{2}((\tilde{y}_k(B))^\intercal S_k^{-1} \tilde{y}_k(B) + \log
  |S_k| + d_y \log 2 \pi),
\end{equation}
where $c$ is a positive regularization parameter and $\Delta$ denotes
the first difference. The second part of the cost function in
\eqref{eq:fullOpt} thus constitutes a regularization term which
reduces high-frequency fluctuations in the estimate. $\tilde{y}_k$ is
the deviation between measurements $z_k$ and the output from the
mean-field dynamics system:
\begin{equation}
  \tilde{y}_k = z_k - H x_k.
\end{equation}
There is no correction of the state from the measurements as in the
Kalman filter, but instead the state $x_k$ is given by the state space
model
\begin{equation}
  x_{k+1}=F_k x_k,
\end{equation}
i.e., the matrix $F_k$ is now time varying per day $k$ according to
\begin{equation}
  F_k = F(\beta_k),
\end{equation}
where the remaining parameters of $F_k$ are given by the previously
inferred posterior maximum likelihoods.  Finally, the covariance
matrices $S_k$ in \eqref{eq:fullOpt} are calculated from the linear
filter, again with the maximum likelihood parameter values, and
including $\beta_t$. As a result, the optimization can be viewed as a
quasi-maximum likelihood estimation. Here, the parameter $\beta_t$ is
estimated, as expected, to show the fastest temporal fluctuations, but
problems with several free parameters could also be considered, such
as the IFR. The initial state $x_0$, could also be included in the
optimization formulation.

The problem is solved using the interior point method with the
function \texttt{fmincon} in Matlab. Including the whole optimization
horizon in a single optimization, however, turns out to be very time
consuming computationally, or even infeasible for the problem at hand,
when $K$ is of the order of several hundreds. For this reason, an
approach inspired by techniques from automatic control is used to
divide the optimization horizon as described below.

\subsection*{Dividing the optimization horizon}

Since the complexity of the optimization algorithm is superlinear in
$K$, computational gains can be made by dividing the optimization
horizon into shorter windows, which are solved independently. In our
context, an additional motivation for this is that the problem is
solved every week as new data becomes available, which means that
optimization results for earlier time windows potentially could be
reused. However, the result of concatenating optimization results
calculated over non-overlapping windows does not coincide with the
solution to the full optimization \eqref{eq:fullOpt}, due to
end-of-horizon effects, i.e., that the $\beta_t$-estimates toward the
end of one time window do take data outside the window into
account. Inspired by the methodology of Model Predictive Control
(MPC), we therefore utilize overlapping optimization windows. More
specifically, time steps $\Delta k = k_{i+1}-k_i$ are used to iterate
over the optimization horizon, corresponding to the sampling times in
MPC, and at each step, an optimization of the form \eqref{eq:fullOpt}
is solved, but over a horizon $[k_i,k_i+K_p]$, where $\Delta
k<K_p<K$. This horizon corresponds to the prediction horizon in
MPC. The calculated values of $\beta_k$ for $k\in [k_i,k_{i+1}]$ are
then used to build the vector $\tilde B$. Assuming that for a
sufficiently large $K_0$, measurements at times $\kappa_0+L, L\ge K_0$
have negligible effect on the values of the optimal
$\beta_k, k\le \kappa_0$, we are thus able to replace the optimization
problem \eqref{eq:fullOpt} with a sequence of optimization problems of
the form
\begin{equation} \tilde B_i = \argmin_{B_i} \sum_{k=k_i}^{k_{i+1}}
  \frac{1}{2}(((\tilde{y}_k(B_i))^\intercal S_k^{-1} \tilde{y}_k(B_i) +
  \log |S_k| + d_y \log 2 \pi) + c \Delta B_i^\intercal \Delta B_i,
\end{equation} where $\tilde B$ then is created by concatenating the
first $\Delta k$ elements of each $\tilde B_i$ (except for the last
$\tilde B_i$ which is used in its entirety). Notice that constraints
need to be added to the optimizations to ensure ``continuity'' of
$\tilde B$, i.e. that the regularization is employed also across the
limits of the time windows.

\begin{SCfigure}
  \includegraphics[clip = true, trim =
   4cm 9.8cm 5.25cm 9.5cm,width=0.5\textwidth] {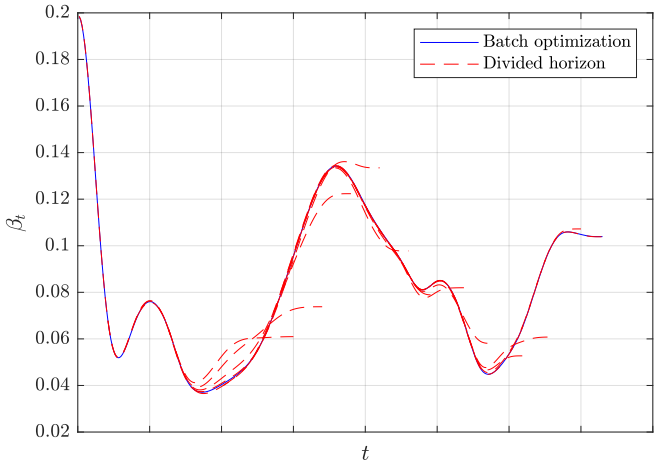}
  \caption{Estimated $\beta_t$ with one batch optimization versus
    dividing the horizon. For the latter method, the discarded
    ``tails'' of each optimization which are caused by the
    end-of-horizon effect are visible.}
  \label{fig:horizonTest}
\end{SCfigure}

In our case, a prediction horizon of $150$ days and a step length of
$20$ days was used. For typical datasets, there is then no discernible
difference between the optimal solution for the full horizon and the
combination of the solutions to the smaller problems. In
Fig.~\ref{fig:horizonTest}, this is illustrated for the estimation of
$\beta_t$ from one year of data from Uppsala. The solution time with
the divided optimization is shorter; approximately seven minutes
instead of ten on a standard modern laptop. This difference increases
with the length of the total time horizon.

\section*{Additional evaluations}

\subsection*{Posterior robustness}

\begin{figure}[tbhp]
  \centering
  \includegraphics[clip = true,
  trim = 1.0cm 7.5cm 1.0cm 7.5cm, 
  width=1\textwidth] {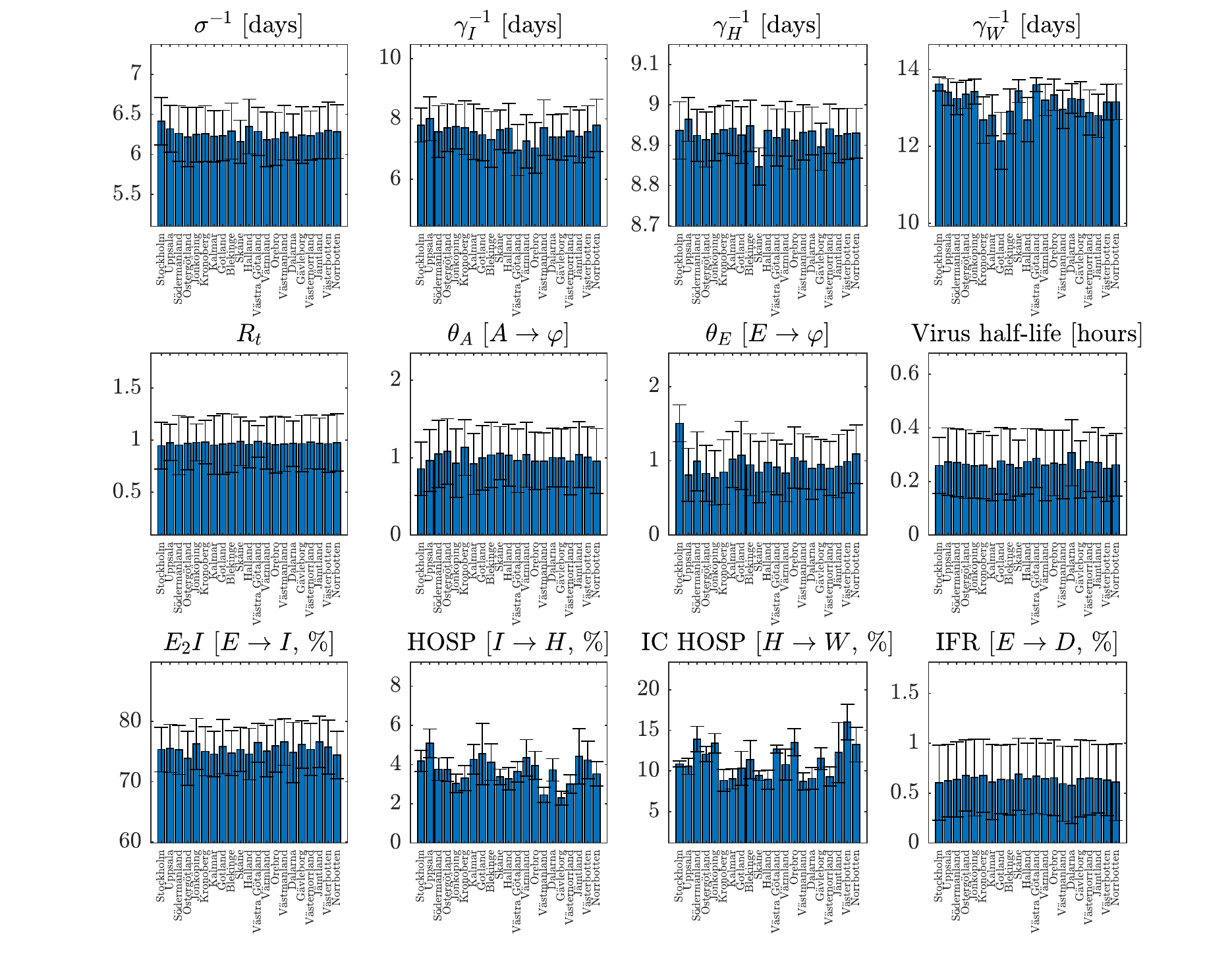}
  \caption{Posterior mean $\pm$ 1 standard deviation per parameter and
    across 21 Swedish regions. For the dynamic parameters $R_t$ and
    IFR, the temporal average is displayed.}
  \label{fig:errorbars_region}
\end{figure}

The computed posterior in Fig.~\ref{fig:posterior_sweden} is the
population weighted posterior when combining the samples from Sweden's
21 regions. In Fig.~\ref{fig:errorbars_region} we display the mean
posterior $\pm$ 1 standard deviation for each region. The results
agree well across this natural bootstrap population albeit with a few
outliers.

\subsection*{Bootstrap robustness}

\begin{figure}  \centering
  \begin{subfigure}{.3\textwidth}
    \includegraphics[clip=true, trim = 4.5cm 9.3cm 5.2cm 9.5cm,
    width=1\linewidth] {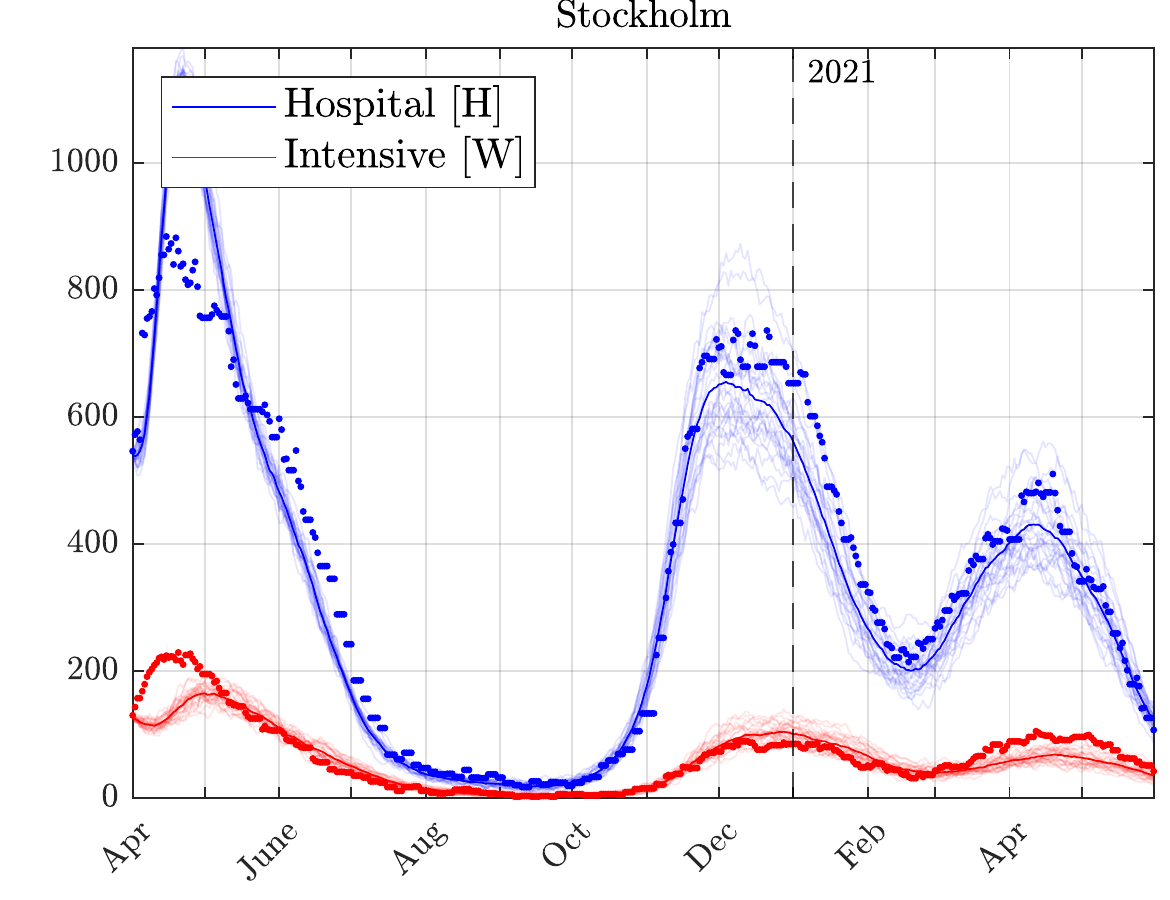}
  \end{subfigure}
  \hfill
  \begin{subfigure}{0.3\textwidth}
    \includegraphics[clip=true, trim = 4.5cm 9.3cm 5.2cm 9.5cm,
    width=1\linewidth]{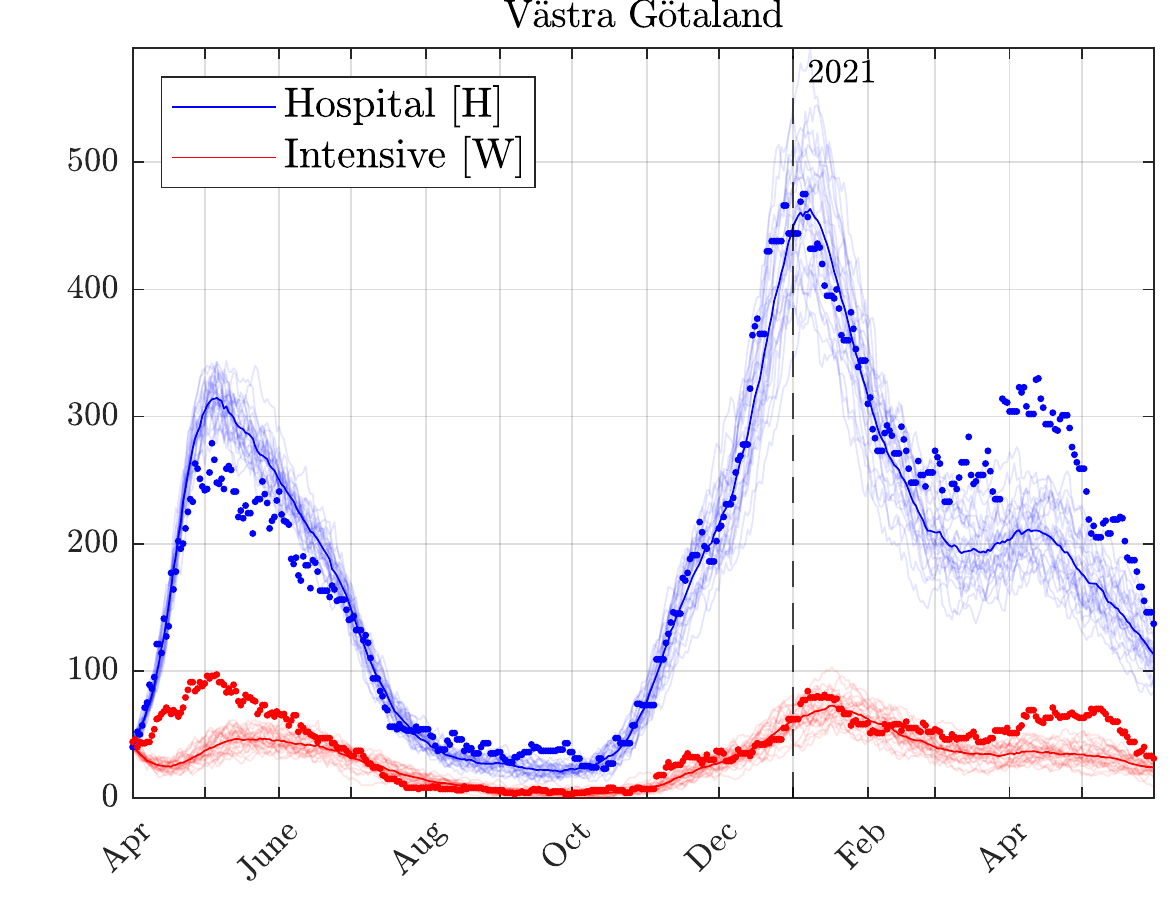}
  \end{subfigure}
  \hfill
  \begin{subfigure}{0.3\textwidth}
    \includegraphics[clip=true, trim = 4.5cm 9.3cm 5.2cm 9.5cm,
    width=1\linewidth]{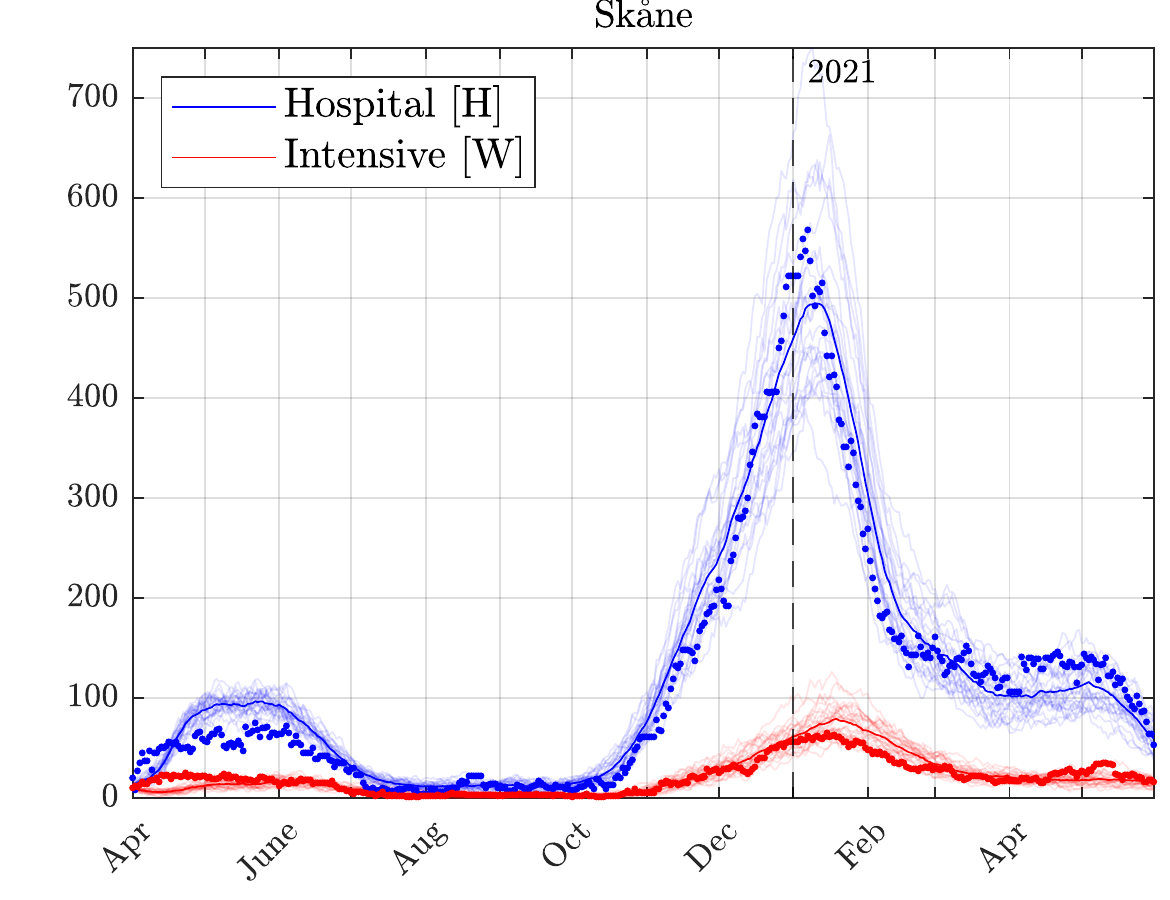}
  \end{subfigure}
  \newline
  \begin{subfigure}{0.3\textwidth}
    \includegraphics[clip=true, trim = 4.5cm 9.3cm 5.2cm 9.5cm,
    width=1\linewidth]{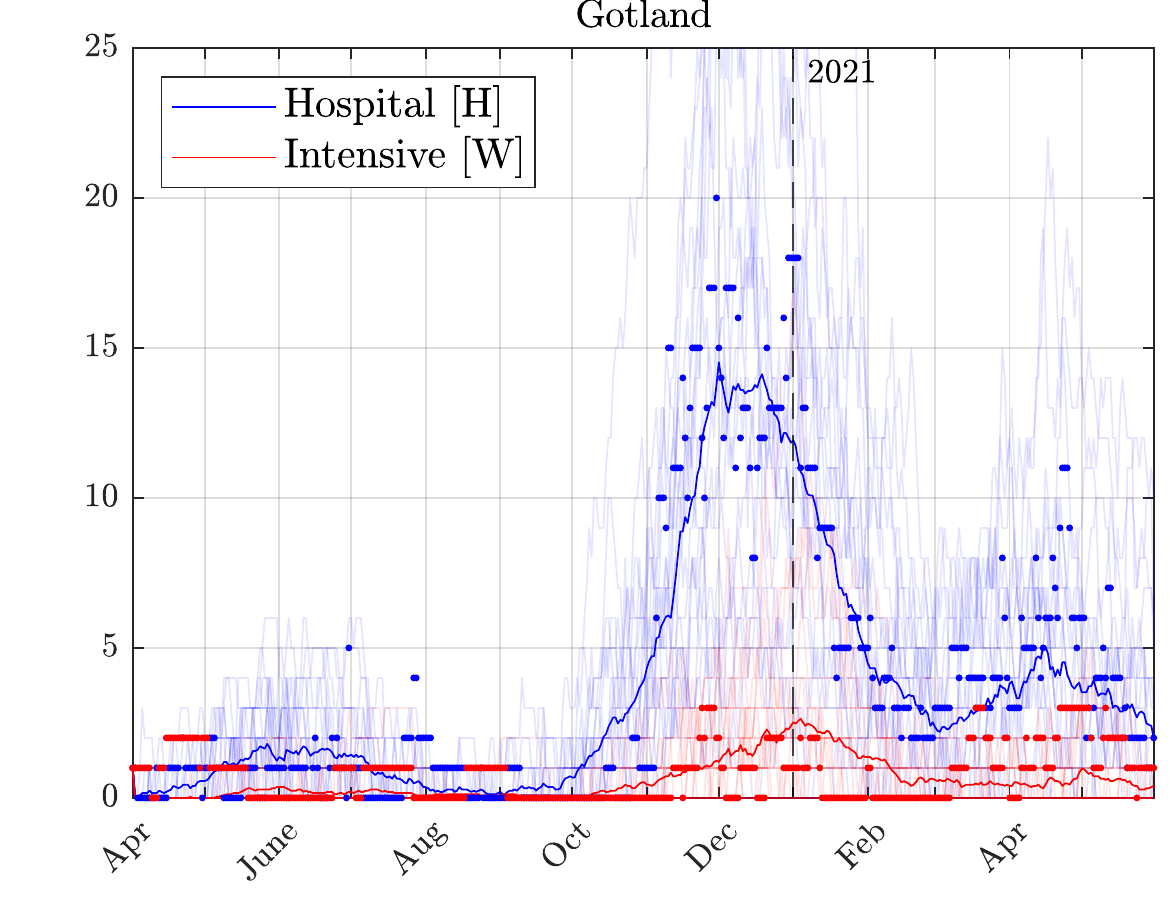}
  \end{subfigure}
  \hfill
  \begin{subfigure}{0.3\textwidth}
    \includegraphics[clip=true, trim = 4.5cm 9.3cm 5.2cm 9.5cm,
    width=1\linewidth]{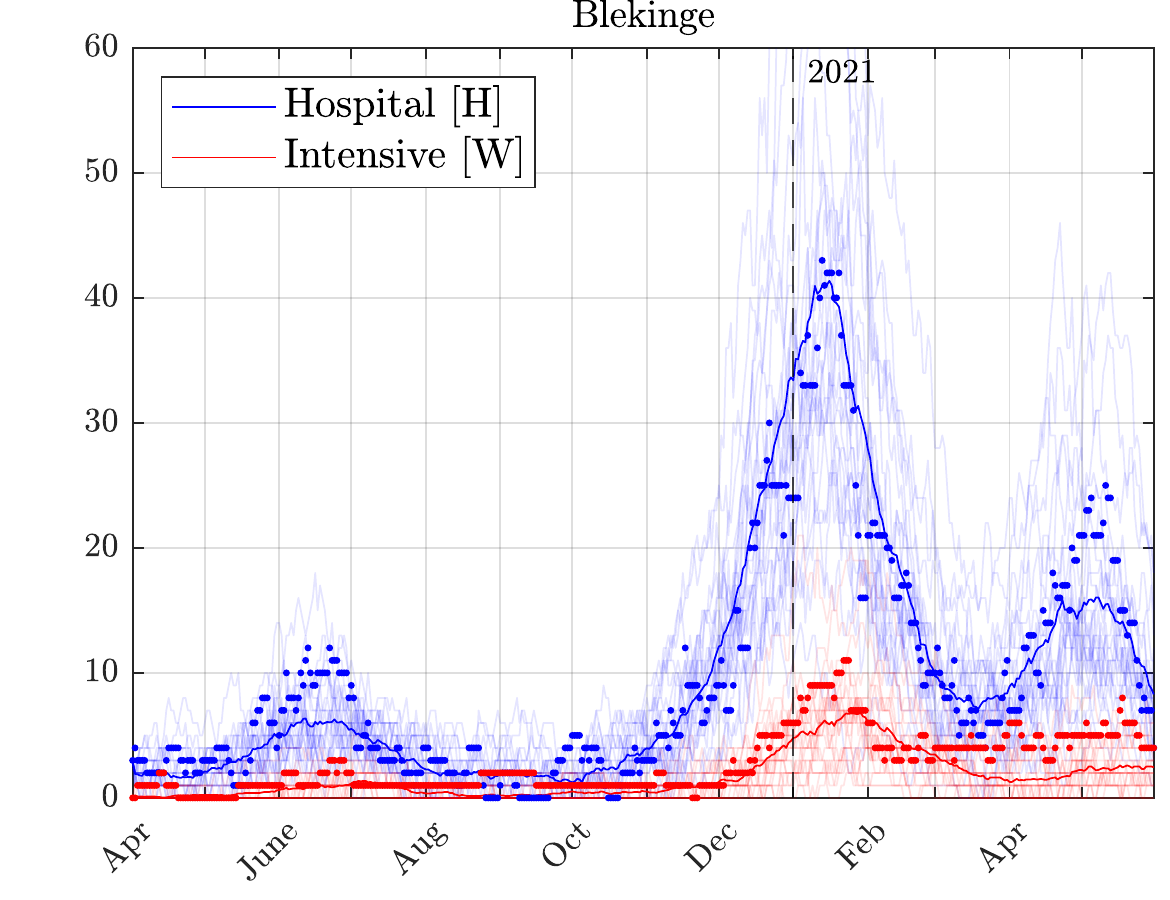}
  \end{subfigure}
  \hfill
  \begin{subfigure}{0.3\textwidth}
    \includegraphics[clip=true, trim = 4.5cm 9.3cm 5.2cm 9.5cm,
    width=1\linewidth]{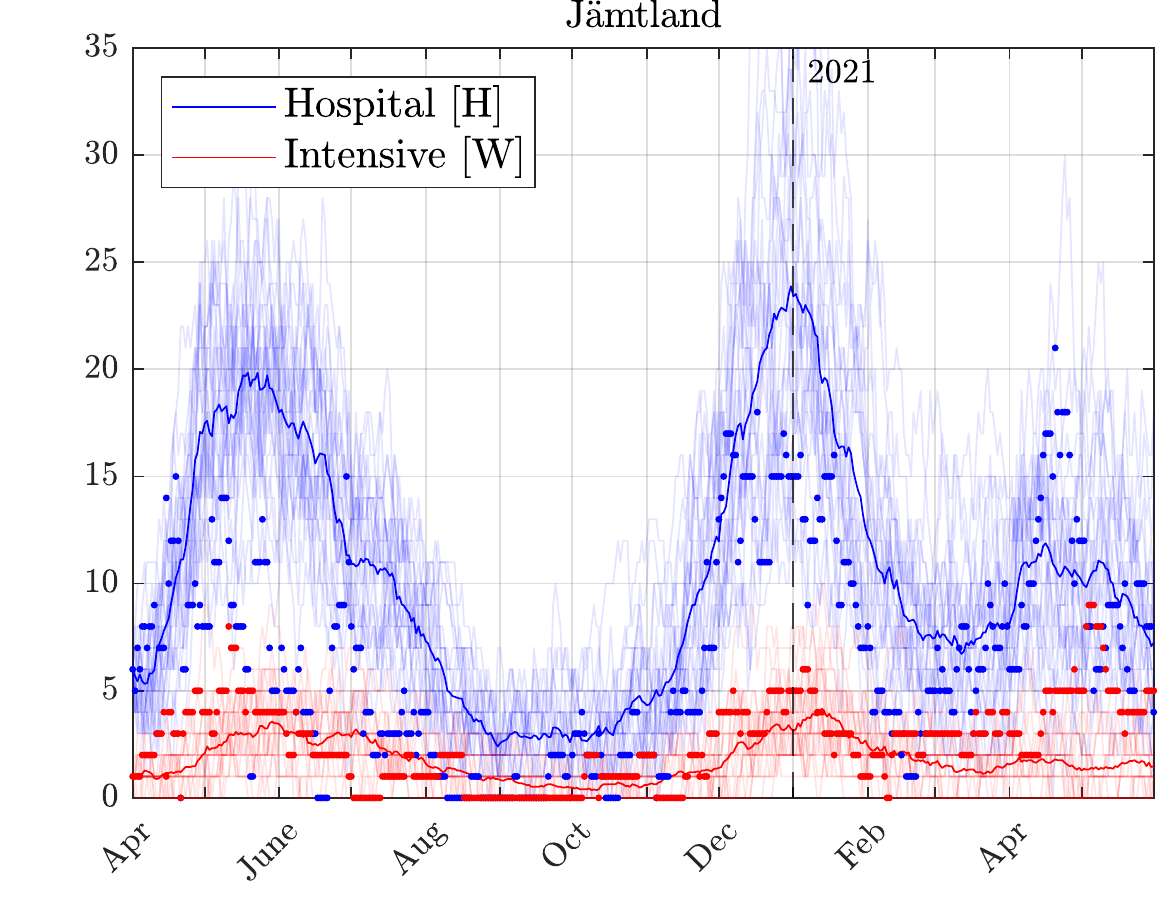}
  \end{subfigure}
  \newline
  \begin{subfigure}{0.3\textwidth}
    \includegraphics[clip=true, trim = 4.5cm 9.3cm 5.2cm 9.5cm,
    width=1\linewidth]{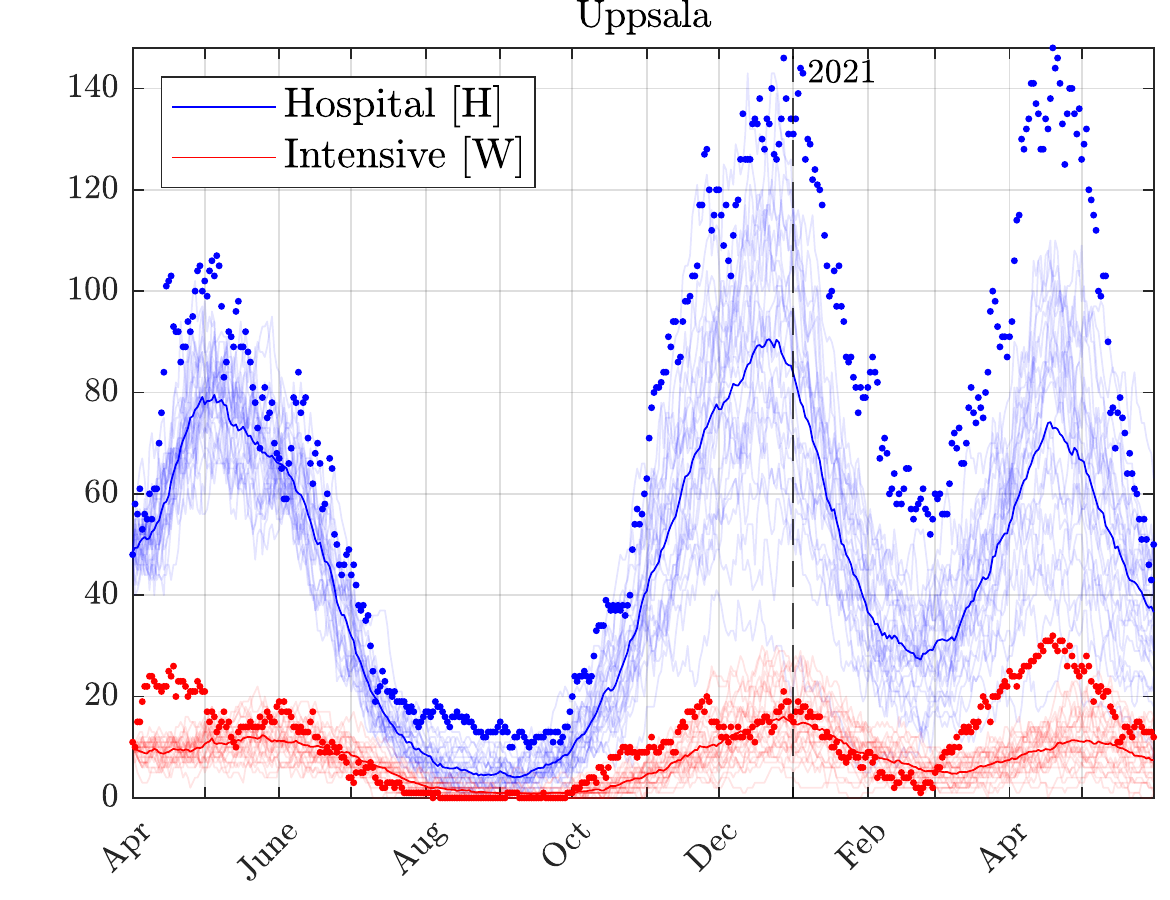}
  \end{subfigure}
  \caption{Fully synthetic simulations for a few selected regions with
    parameters from the national mean posterior, but with upscaled
    regional $\beta_t$. The lines of lighter shades of blue and red
    are realizations and the solid lines are the mean of the samples.
    The points are the data points used in the inference.}
  \label{fig:URDME_samples}
\end{figure}

By the bootstrap procedure presented in \matmet, we can investigate the
posterior robustness, including estimating the bias due to the
approximate likelihood. The inference procedure is repeated on a
synthetic data set generated by the mean posterior estimate and with
the temporal resolution upscaling procedure for $\beta_t$. In
Fig.~\ref{fig:URDME_samples} we display 25 such replicates for a few
selected regions. Note that these were obtained in a completely
off-line fashion and are never corrected against data.

For each region $r = 1,\dots,21$ and parameter dimension $k=1,\dots,K$
in the posterior we estimate the bias $\tilde{b}_{r,k}$, as per the
description in \matmet. The dynamic parameters are here treated as
a single parameter, that is, with a single average bias, and together
with the static parameters there are $K = 12$ in total. We compute
three statistics $T = [T_1, T_2, T_3]$ to characterize the spread:
the coefficient of variation (CoV), the coefficient of bias (CoB), and
the normalized root-mean-square error (NRMSE),
\begin{equation}\label{eq:bootstat}
  T_{k,r,1} = \text{CoV} =\sigma_{k,r}/\mu_{k,r}, \quad
  T_{k,r,2} = \text{CoB}= |\tilde{b}_{k,r}|/\mu_{k,r},\quad
  T_{k,r,3} =  \text{NRMSE}  = \left(\sigma^2_{k,r} +
      \tilde{b}^2_{k,r}\right)^{1/2}/\mu_{k,r},
\end{equation}
for the standard deviation $\sigma_{k,r}$, the mean $\mu_{k,r}$, and
the estimated bias $\tilde{b}_{k,r}$. In Tab.~\ref{tab:bootstrap}, we
present the median value per region and statistic $s \in \{1,2,3\}$
over the $K$ parameters,
\begin{equation}\label{eq:mean_bootstat}
  \bar{T}_{r,s} = \text{Median}_k \, T_{k,r,s}.
\end{equation}

\begin{figure}[tbhp]
  \centering
  \includegraphics[clip = true, trim =  2.0cm 8.2cm 2.0cm 8.0cm]
  {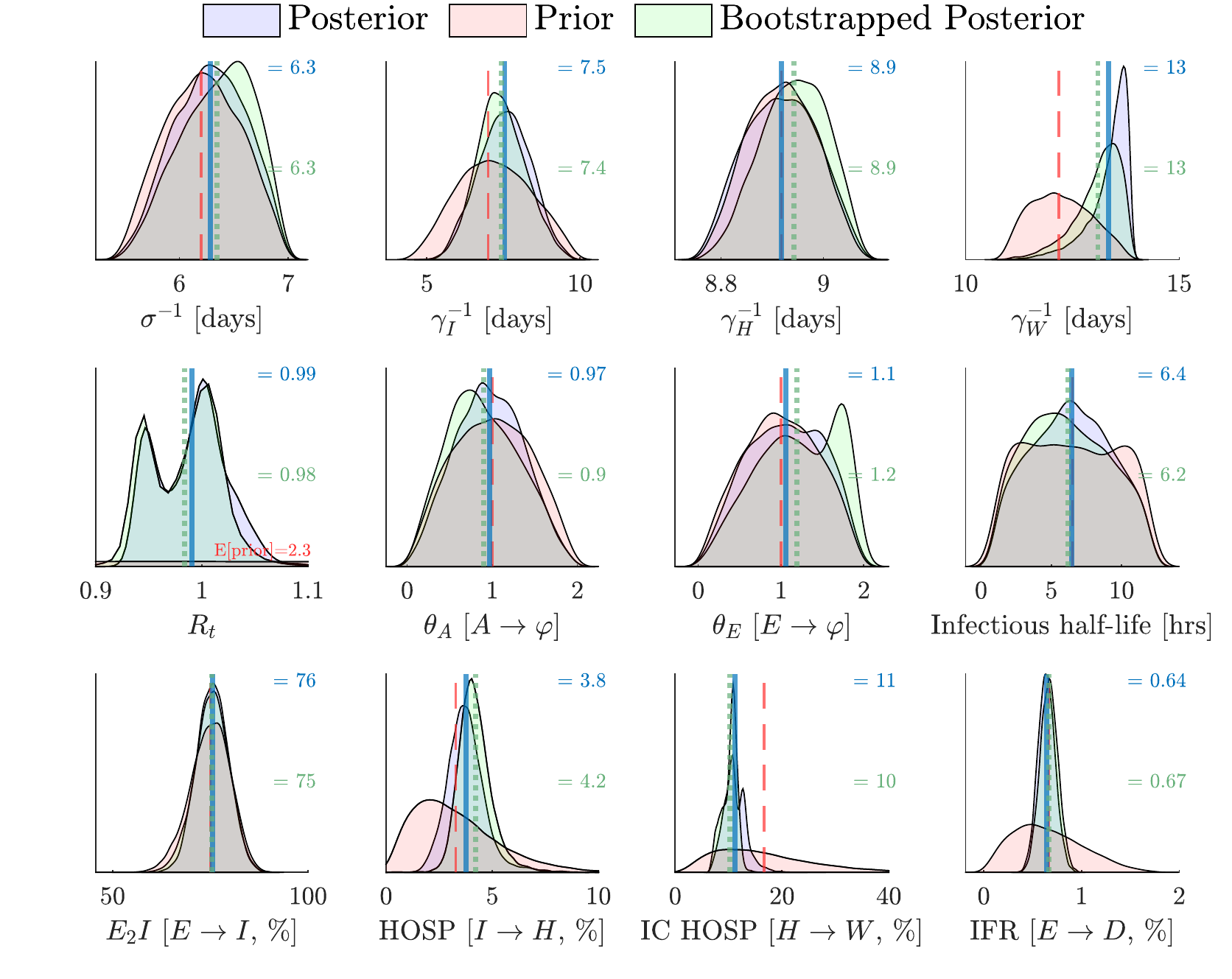}
  \caption{Population weighted national average posterior using:
    actual data (blue) or bootstrapped data (green), for a selected
    set of parameters. Both posteriors use the same prior (red). The
    vertical lines indicate the density means: posterior (blue solid),
    prior (red dashed), and bootstrapped posterior (green dotted). The
    mean is also annotated in the corresponding color for the
    posterior estimates.}
  \label{fig:posterior_urdmesweden}
\end{figure}

\begin{figure}[tbhp]
  \centering
  \includegraphics[clip = true, trim = 2.5cm 9.2cm 2.8cm 8.5cm]
  {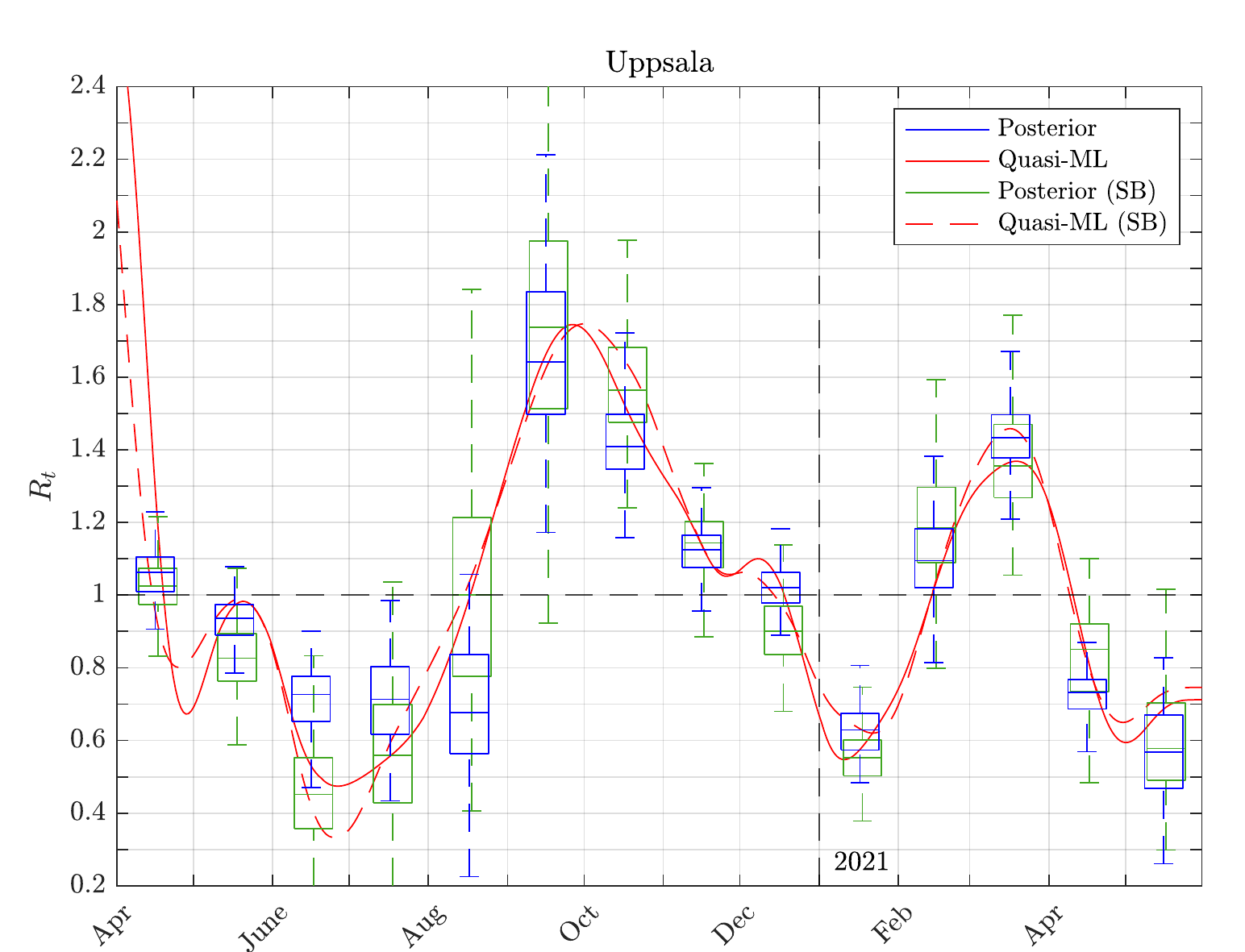}
  \caption{Reproduction number estimates for the Uppsala region with
    bootstrap replicas. The Bayesian posterior on the true data yields
    monthly estimates (blue box-plot) while the quasi-ML estimator is
    daily (red solid). The aggregate posterior of the bootstrap
    replicates $[n=3]$ (green box-plot) and the average daily quasi-ML
    estimator (red dashed) from the bootstrap replicates.}
  \label{fig:Rposterior_urdme}
\end{figure}

We visualize both the data posterior and the aggregate of all
bootstrap replicate posteriors ($n=3$) in
Fig.~\ref{fig:posterior_urdmesweden}. We also compare the daily $R_t$
estimate of the actual data posterior and the average from the
bootstrap replicates for Uppsala in Fig.~\ref{fig:Rposterior_urdme}.

\begin{table}[htp]
  \centering
  \begin{tabular}{lrrr|rr}
    \hline
    Region & CoV [\%] & CoB [\%] & NRMSE [\%] &
                                                $d_{\text{smooth}} [\%]$ & Population 	\\
    \hline
    Stockholm & 6.3 & 4.9 & 12 & 6.6 & $2.4 \cdot 10^{6}$ 	\\
    Uppsala & 13 & 2.6 & 15 & 6.6 & $3.8 \cdot 10^{5}$ 	\\
    Sodermanland & 15 & 2.0 & 18 & 8.1 & $3.0 \cdot 10^{5}$ 	\\
    Ostergotland & 13 & 4.3 & 19 & 7.3 & $4.7 \cdot 10^{5}$ 	\\
    Jonkoping & 14 & 2.4 & 18 & 4.7 & $3.6 \cdot 10^{5}$ 	\\
    Kronoberg & 14 & 4.0 & 24 & 6.9 & $2.0 \cdot 10^{5}$ 	\\
    Kalmar & 16 & 3.4 & 23 & 5.1 & $2.5 \cdot 10^{5}$ 	\\
    Gotland & 29 & 3.2 & 32 & 2.2 & $6.0 \cdot 10^{4}$ 	\\
    Blekinge & 18 & 3.1 & 30 & 13 & $1.6 \cdot 10^{5}$ 	\\
    Skane & 12 & 3.6 & 13 & 5.9 & $1.4 \cdot 10^{6}$ 	\\
    Halland & 16 & 2.4 & 24 & 9.6 & $3.3 \cdot 10^{5}$ 	\\
    Vastra Gotaland & 11 & 3.9 & 14 & 6.2 & $1.7 \cdot 10^{6}$ 	\\
    Varmland & 18 & 3.5 & 25 & 7.7 & $2.8 \cdot 10^{5}$ 	\\
    Orebro & 14 & 2.6 & 20 & 9.4 & $3.0 \cdot 10^{5}$ 	\\
    Vastmanland & 15 & 5.4 & 24 & 4.4 & $2.8 \cdot 10^{5}$ 	\\
    Dalarna & 14 & 3.6 & 24 & 5.7 & $2.9 \cdot 10^{5}$ 	\\
    Gavleborg & 14 & 2.7 & 22 & 4.8 & $2.9 \cdot 10^{5}$ 	\\
    Vasternorrland & 14 & 2.8 & 24 & 4.3 & $2.5 \cdot 10^{5}$ 	\\
    Jamtland & 22 & 4.1 & 28 & 13 & $1.3 \cdot 10^{5}$ 	\\
    Vasterbotten & 20 & 2.0 & 23 & 10 & $2.7 \cdot 10^{5}$ 	\\
    Norrbotten & 18 & 2.6 & 26 & 4.5 & $2.5 \cdot 10^{5}$ 	\\
    \hline
  \end{tabular}
  \caption{Median uncertainty statistic per region: CoV, CoB, and NRMSE
    as in Eqs.~\eqref{eq:bootstat}--\eqref{eq:mean_bootstat}. The
    smoothing difference $d_{{\normalfont \text{smooth}}}$ is the mean
    max relative difference between the pre-processed and raw data
    $[H, W, D]$ as defined in \eqref{eq:minmaxdiff}.}
  \label{tab:bootstrap}
\end{table}

\subsection*{Baseline predictor}

To evaluate the predictive performance of our Kalman filter model
outside of a model vacuum, we compare to an autoregressive model (AR)
model. AR models are common in time-series predictions, and for
COVID-19, AR models with auxiliary indicators show significant
prediction power \cite{mcdonald2021can}.  We consider a single-day
forward expanding data window for which we fit the AR model on all
data in the window and predict 7 days ahead. This procedure is
evaluated across the same data as used in our live reports ($N = 25$).

As the Kalman filter, the AR model also uses $[H,W,D]$ as
observations. We use Matlab's \verb;arx; implementation of the AR
model, formally called Vector AR model with Exogenous Variables. The
polynomial orders and delays are set from finding good predictions in
a mean square error sense on the first 50 days of the dataset. We tune
$n_b$ and $n_c$ by hand, and proposed values for $n_a$ by a partial
autocorrelation function plot per data dimensions.  We find the order
of the $A_q$ polynomials: $n_a = [1, 1, 1]^\intercal$, the $B_q$
polynomials: $n_b = [0, 1, 0]^\intercal$, and the input-output delay:
$n_c = [1, 1, 1]^\intercal$ to be close to optimal choices.

\begin{figure}[tbhp]
  \centering
  \includegraphics[clip = true,
  trim = 4.5cm 9.3cm 4.5cm 9.5cm]{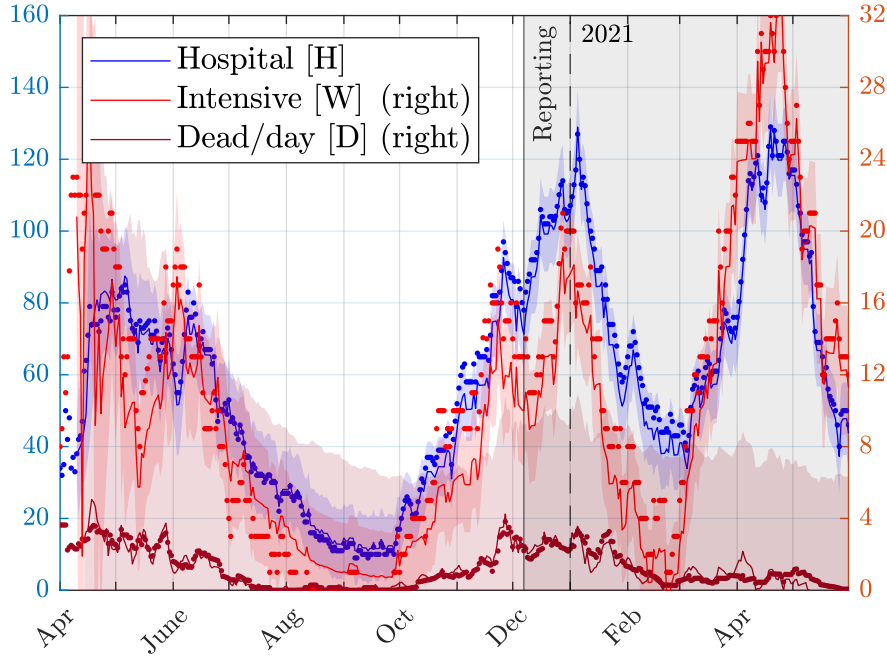}
  \caption{Result from the 7-day ahead AR prediction for the Uppsala
    region with 68\% CrI (shaded). The live reporting ($N = 25$) for
    Tabs.~\ref{tab:PastWeekly} and \ref{tab:modelcomparison} was done
    in the reporting period indicated towards the second half.}
  \label{fig:modelcomparison}
\end{figure}

In Fig.~\ref{fig:modelcomparison}, we visualize the 7-day ahead
prediction made by the AR model for Uppsala. In
Tab.~\ref{tab:modelcomparison}, we present the respective prediction
precision by NRMSE and multivariate Energy score
\cite{gneiting2008assessing} of the two models along with a repetition
of the results from Tab.~\ref{tab:PastWeekly}. A closer inspection
reveals that the simpler models have a smaller NRMSE and Energy Score
than the posterior Kalman filter but with overly pessimistic CrIs. The
simple AR model generates good mean-square predictions in a fraction
of the training time of our posterior Kalman filter and could likely
be improved a bit when it comes to the width of the confidence
interval estimate, reducing the Energy score further. The great
advantage with our approach lies instead in the fact that the
posterior model itself can be disassembled and contains valuable
epidemiological information.

\begin{table}[htp]
  \centering
  \begin{tabular}{lrrr}\hline
    & Hospital (H)& Intensive (W)& Death (D)\\\hline
    Posterior Kalman (68\% CrI)& 76& 72& 68\\
    AR& 100& 100& 44\\\hline
    Posterior Kalman (95\% CrI)& 100& 100& 96\\
    AR& 100& 100& 92\\\hline
    Posterior Kalman (NRMSE)& 25& 27& 1.9\\
    AR& 4.7& 12& 1.4\\\hline
    Posterior Kalman (Energy score)& 84& 15& 30\\
    AR& 18& 7.1& 24\\
    \hline
  \end{tabular}
  \caption{Frequency of all weekly reported predictions (N~=~25, Dec
    2020--May 2021) for Uppsala that fell inside of the reported CrIs
    (68/95\%, 7 days ahead), the NRMSE, and the multivariate Energy
    Score evaluated on the following week, thus comparing the
    performance of the posterior Kalman filter and the AR model.}
  \label{tab:modelcomparison}
\end{table}



\end{document}